\documentclass[a4paper]{JHEP3}
\usepackage[centertags]{amsmath}
\usepackage{amssymb}
\usepackage{fixltx2e}
\MakeRobust{\eqref} 
\usepackage{graphicx}
\usepackage{subfigure}
\preprint{TIFR/TH/13-17}
\newcommand{\half}{\frac{1}{2}}
\newcommand{\Tr}{\text{Tr}}

\def\cC{{\cal C}}

\def\cD{{\cal D}}

\def\cS{{\cal S}}

\def \beal#1 {\begin{align}#1\end{align}}

\def\bS{{\mathbf S}}
\def\bR{{\mathbf R}}

\def\tr{\mathrm{tr}}
\def\Tr{\mathrm{Tr}}

\def\half{{1\over2}}
\def\nn{\nonumber\\}
\def\sgn {\text{sgn}}

\def\[{\left[}
\def\]{\right]}
\def\({\left(}
\def\){\right)}

\newcommand{\wt}{\widetilde}

\def\={\stackrel{\bullet}{=}}
\def\nn{\notag\\}

\def\Tr{\mathrm{Tr}}

\def\half{{1\over2}}

\def\[{\left[}
\def\]{\right]}
\def\({\left(}
\def\){\right)}

\def\cD{{\cal D}}

\def\cS{{\cal S}}

\def\bS{{\mathbf S}}

\def\bR{{\mathbf R}}

\def \be {\begin{equation}}
\def \ee {\end{equation}}
\def \bea {\begin{eqnarray}}
\def \eea {\end{eqnarray}}
\def \beal#1 {\begin{align}#1\end{align}}
\def \nn {\notag\\}

\def\aver#1{\left\langle #1 \right\rangle}

\newcommand{\lm}{\lambda}

\def \be {\begin{equation}}
\def \ee {\end{equation}}
\def \bea {\begin{eqnarray}}
\def \eea {\end{eqnarray}}
\def \nn {\notag\\}

\def\aver#1{\left\langle #1 \right\rangle}

\usepackage{enumerate}

\title{Chern Simons duality with a fundamental boson and fermion}

\author{
Sachin Jain$^{1}$, 
Shiraz Minwalla$^{2}$, 
Shuichi Yokoyama$^{3}$
\\
Department of Theoretical Physics, Tata Institute of Fundamental Research,\\
Homi Bhabha Road, Mumbai 400005, India\\
{\small \tt E-mail: $^1$sachin, ${}^2$minwalla, ${}^3$yokoyama(at)theory.tifr.res.in}
}
\abstract{We compute the thermal free energy 
for all renormalizable Chern Simon theories coupled to 
a single fundamental bosonic and fermionic field in the 't Hooft 
large $N$ limit. We use our results to 
conjecture a strong weak coupling duality invariance for this class of
theories. Our conjectured duality reduces to Giveon Kutasov duality 
when restricted to ${\cal N}=2$ supersymmetric theories and to an earlier 
conjectured bosonization duality in an appropriate decoupling limit. 
Consequently the bosonization duality may be regarded 
as a deformation of Giveon Kutasov duality, suggesting that it 
is true even at large but finite $N$.}

\begin{document}

%%%%%%%%%%%%%%%%%%%%%%%%%%%%%%%%%%%%%%%%%
\section{Introduction}

Three dimensional Chern Simons $U(N)$ gauge theories coupled to matter fields
are interesting from several points of view. These theories have applications 
as diverse as quantum hall physics \cite{Frohlich:1990xz,Frohlich:1991wb}, the topology of three manifolds \cite{Witten:1988hf}
and the study of quantum gravity via the AdS/CFT correspondence \cite{Aharony:2008ug}. 

It has recently been realized that vector Chern Simons theories (i.e. 
Chern Simons theories whose matter fields are all in the fundamental of $U(N)$) 
are `solvable' in the large $N$ limit. 
We now have explicit formulae for the spectrum of single sum operators, 
three point correlation functions and thermal partition functions for  
some of these theories 
\cite{Giombi:2011kc,Aharony:2011jz,Maldacena:2011jn,Maldacena:2012sf,Aharony:2012nh,Jain:2012qi,Yokoyama:2012fa,Giombi:2012ms,Aharony:2012ns,
GurAri:2012is,Jain:2013py,Takimi:2013zca}. These exact results have  already
led to the discovery of new non supersymmetric level rank dualities  between 
classes of such theories. The existence of such a duality was 
first speculated on in \cite{Giombi:2011kc}; concrete evidence for such a
duality was provided by the study of three point functions 
in \cite{Maldacena:2011jn,Maldacena:2012sf}. The transformation of parameters 
under duality was first worked out in \cite{Aharony:2012nh} (based on 
correlation function computations, see also \cite{GurAri:2012is}); 
this paper also made the first definite conjecture 
for such a duality. The equality of free energies of dual pairs was 
established in \cite{Aharony:2012ns,Jain:2013py,Takimi:2013zca}.

Vector matter Chern Simons theories have also been conjectured to admit dual 
bulk descriptions as a theory of higher spin fields \cite{Giombi:2011kc,Aharony:2011jz,Chang:2012kt}. It thus appears 
that these theories are dynamically rich even in the analytically 
tractable large $N$ limit.

In this note we study the most general renormalizable $U(N)$ Chern Simons 
theory coupled to a single fundamental scalar and fermion field in the large 
$N$ limit. In particular we use the methods of 
\cite{Giombi:2011kc,Jain:2012qi,Aharony:2012ns,Jain:2013py,Yokoyama:2012fa} to exactly determine the quantum corrected pole masses 
of the propagating bosonic and fermionic fields in this theory. 
We also determine the exact thermal 
partition function of these theories as a function of temperature, chemical 
potentials and couplings. We use our results to propose a level rank type 
strong weak 
coupling duality map that relates pairs of these theories, 
generalizing an earlier proposal by \cite{Aharony:2012ns}.

A special case of the theories we study in this paper is 
the superconformal level $k$ $U(N)$ ${\cal N}=2$ Chern Simons 
theory with a single fundamental chiral multiplet. The action for this 
theory is given in terms of component 
fields by\footnote{The conventions we have employed in that action are 
\beal{
\wt D_{\mu} \phi =&(\partial _{\mu} -i A_\mu - \mu_B \delta_{3\mu} ) \phi \quad 
\wt D_{\mu} \bar \phi =\partial _{\mu} \bar \phi+ i \bar \phi A_\mu + \mu_B \delta_{3\mu}\bar \phi  \\
\wt D_{\mu} \psi =&(\partial _{\mu} -i A_\mu - \mu_F \delta_{3\mu} ) \psi \quad 
\wt D_{\mu} \bar \psi =\partial _{\mu} \bar \psi + i\bar \psi A_\mu + \mu_F \delta_{3\mu} \bar \psi.
}
}   
\beal{
S  &= \int d^3 x  \biggl[i \varepsilon^{\mu\nu\rho} {k \over 4 \pi}
\Tr( A_\mu\partial_\nu A_\rho -{2 i\over3}  A_\mu A_\nu A_\rho)
+ \wt D_\mu \bar \phi \wt D^\mu\phi + \bar\psi \gamma^\mu \wt D_\mu \psi \nn
&
+ \frac{4 \pi}{k} (\bar\psi \psi) (\bar\phi\phi)
+ \frac{2 \pi}{k} (\bar\psi\phi)( \bar\phi \psi)
+ \frac{4 \pi^2}{k^2} (\bar\phi\phi)^3  \biggl]
\label{susyaction}
}
where $\phi$ and $\psi$ respectively are complex scalar and spinor
fields that transform in the fundamental representation of $U(N)$.
The action \eqref{susyaction} defines\footnote{More precisely, the field theories 
we study are defined by their action together with a regulation scheme. 
Throughout this paper we employ the dimensional reduction scheme employed 
in \cite{Giombi:2011kc,Jain:2012qi}. 
In this regularization scheme the Chern-Simons level $k$ satisfies 
$|k|= |k_{\text{YM}}|+N$, where $k_{\text{YM}}$ is the Chern-Simons level in 
the regularization by the Yang-Mills term \cite{Pisarski:1985yj,Chen:1992ee}. 
This requires the 't Hooft coupling defined by $\lambda=N/k$ to satisfy $|\lambda|\leq1$.
Through this paper we also work in the lightcone gauge of \cite{Giombi:2011kc}.}
 a discrete set of superconformal ${\cal N}=2$ field 
theories labeled by the two integers $N$ and $k$. There is considerable 
evidence that the theories 
\eqref{susyaction} enjoy a strong weak coupling level rank type duality 
under which $k$ goes to $ -k$ and $N$ goes to $|k|-N$; recent 
evidence for this supersymmetric level rank duality includes the matching of 
the thermal partition function of conjecturally dual pairs of theories 
in the large $N$ limit \cite{Aharony:2012ns,Jain:2013py,Takimi:2013zca}.

The duality between pairs of  ${\cal N}=2$ theories \eqref{susyaction} 
implies a duality map on the manifold of
 quantum field theories obtained  by perturbing 
\eqref{susyaction} with relevant or marginal operators. 
In the large $N$ limit relevant deformations of the ${\cal N}=2$ theory are 
mass terms for $\phi$ and $\psi$ plus a  $(\bar \phi \phi)^2$ 
term; marginal deformations of the theory consist of a $({\bar \phi} \phi)^3$ 
potential together with  3 distinct Yukawa terms. The manifold of quantum 
field theories obtained by perturbing the ${\cal N}=2$ field theory with 
a marginal or relevant operator is spanned by the theories%
\footnote{The operators parameterized by the  deformations $x_6$, $x_4$, 
$y_4'$ and $y_4''$ are marginal only in the strict large $N$ limit. 
At first nontrivial order in the $\frac{1}{N}$ expansion the scaling 
dimensions of the corresponding operators are presumably all renormalized, 
in which case  $r$ out of these operators will turn relevant while the 
remaining $4-r$ operators become irrelevant. The self duality of the 
${\cal N}=2$ theory would then imply the duality of the $3+r$ dimensional 
manifold of theories obtained from relevant deformations of the ${\cal N}=2$ 
superconformal fixed point. We leave further analysis of the finite 
$N$ system, including the determination of the value of the integer $r$, 
to future work.}
\beal{
S  &= \int d^3 x  \biggl[i \varepsilon^{\mu\nu\rho} {k \over 4 \pi}
\Tr( A_\mu\partial_\nu A_\rho -{2 i\over3}  A_\mu A_\nu A_\rho) \nn
&+ \wt D_\mu \bar \phi \wt D^\mu\phi + \bar\psi \gamma^\mu \wt D_\mu \psi 
+m_B^2 \bar\phi \phi +m_F \bar\psi \psi+ {4\pi b_4 \over k} (\bar\phi \phi)^2 
+ \frac{4 \pi^2 x_6}{k^2} (\bar\phi\phi)^3 \nn
&
+ \frac{4 \pi x_4}{k} (\bar\psi \psi) (\bar\phi\phi)
+ \frac{2 \pi y_4'}{k} (\bar\psi\phi)( \bar\phi \psi)
+ \frac{2 \pi y_4''}{k} \left((\bar\psi \phi)( \bar \psi \phi ) 
+(\bar \phi \psi)( \bar \phi \psi )\right) \biggl].
\label{generalaction}
}

In this paper we present exact results for the pole masses and thermal 
free energies of the class of theories \eqref{generalaction} in the 't Hooft large $N$ limit. 
Our results support the conjectured existence of a duality map between these theories, 
and we are able to conjecture a detailed mapping of parameters between pairs 
of (conjecturally) dual theories. Our results generalize the 
results of \cite{Aharony:2012ns} and reduce to those of that paper in the 
special when we set all dimensionful parameters ($m_B$, $m_F$ and $b_4$) 
in \eqref{generalaction} to zero. 

In the rest of this introduction we turn to a more detailed presentation 
of our results and their implications. 
As mentioned above, in this paper we have computed quantum corrected
(zero temperature) pole  masses of the $\phi$ and $\psi$ fields 
respectively. We find that the bosonic pole mass is given by the solution 
to the equation 
\beal{
c^2_{B,0} =& \lambda^2 (1+3x_6) { c^2_{B,0}  \over 4} -2\lambda  b_4 |c_{B,0}|  
 +  x_4 { \lambda (-\lambda + 2 ~\sgn(\rm{X_0})) \over  (\lambda -\sgn(\rm{X_0}) )^2} ( - m_F + x_4 \lambda |c_{B,0}| )^2 + m_B^2 \nn
\label{ztm}
}
and that fermionic pole mass, $c_{F, 0}$, is given in terms of $c_{B,0}$ by 
\begin{equation} \label{fpm}
c^2_{F,0}=\left( {m_F - x_4 |c_{B,0}| \lambda  \over -\lambda + \sgn(\rm{X_0})} \right)^2
\end{equation}
where ${\rm X_0} =\lm(|c_{F,0}|- x_4 |c_{B,0}|)+m_{F}.$

As we demonstrate in the next section, the duality transformation 
\beal{\label{dmap}
& N'=|k|-N, ~~~k'=-k~~~x_4' = {1\over x_4}, \quad  m_F'= -{  m_F \over x_4}\nn
&x_6'=1+{1-x_6 \over x_4^3}, \quad  b_4'=-{1\over x_4^2}( b_4 + {3\over 4} {1-x_6 \over x_4}  m_F)\\
& m_B'^2=- {1\over x_4}  m_B^2 + {3\over 4} {1-x_6 \over x_4^3}  m_F^2 
+ {2 \over x_4^2}  b_4  m_F \notag
}
maps the pair of equations \eqref{ztm} and \eqref{fpm} back to themselves, 
provided we simultaneously interchange the pole masses $c_{B,0}$ and $c_{F,0}$.
Note the 't Hooft coupling $\lambda={N \over k}$ transforms under duality transformation 
as
\be
\lambda'=\lambda - \sgn(\lambda).
\ee 
We also demonstrate invariance of the finite temperature partition
function of these theories under \eqref{dmap} once fermionic and bosonic chemical potentials
are interchanged and the holonomy density function transforms suitably 
(see \eqref{rhotransform}).\footnote{More precisely, as we explain below, the 
gap equations and thermal free energy presented in this paper 
enjoy invariance under duality only when the parameters in 
\eqref{generalaction} obey certain inequalities. See below for more details.}
Motivated by these results we conjecture that 
\eqref{dmap} describes a duality of the quantum theory defined by the Lagrangian 
\eqref{generalaction} in the large $N$ limit. This duality transformation 
interchanges bosons 
and fermions, and so is a `bosonization' duality.\footnote{The submanifold 
on which all dimensionful parameters are set to zero, $m_B=m_F=b_4=0$ 
is preserved by the duality map \eqref{dmap}. Duality also preserves the  
submanifolds $b_4=m_F=0$. However the submanifold $b_4=0$ is not preserved
by duality. }

It is not difficult to verify that \eqref{dmap} is a $Z_2$ duality, i.e. 
that the duality map squares to identity.  

Notice that \eqref{dmap} does {\it not} include a conjecture for the 
transformation rules for $y_4'$ and $y_4''$. The reason for this is 
that the results obtained in this paper turn out to be insensitive
to the values of $y_4'$ and $y_4''$ at leading order in the large $N$ limit. 
The determination of the transformation of the parameters $y_4'$ and $y_4''$
would require either  the evaluation of the partition function at 
a subleading order in the large $N$ expansion or the evaluation 
some other quantity sensitive to these parameters  at leading order in large 
$N$ (this quantity could be  a correlator involving the operator ${\bar \psi} 
\phi$ and ${\bar \phi}\psi$). We leave this interesting task to future work.
We also note that the duality map \eqref{dmap} reduces to that proposed in 
\cite{Aharony:2012ns} in the special case $m_B=m_F=b_4=0$.

Setting 
\begin{equation}\label{susyvalues}
m_F=m_B=b_4=y_4''=0 ~~~x_4=x_6=y_4'=1
\end{equation}
reduces \eqref{generalaction} to the ${\cal N}=2$ supersymmetric 
action \eqref{susyaction}.  It is easily verified that the special 
choice of parameters \eqref{susyvalues} is left unchanged by the duality 
transformation \eqref{dmap}. In other words our conjectured duality \eqref{dmap}
reduces to Giveon Kutasov duality \cite{Giveon:2008zn,Benini:2011mf} ($ N \rightarrow |k|-N$, $k \rightarrow  -k$)
when we make the choice of parameters \eqref{susyvalues}.

Studying \eqref{dmap} in the the neighborhood of the ${\cal N}=2$ point 
\eqref{susyvalues} allows us to deduce the duality transformation rules 
for the relevant and marginal operators of the 
${\cal N}=2$ under Giveon Kutasov duality. Setting 
$m_F= \delta m_F$, $m_B= \delta m_B$, $b_4=\delta b_4$, $x_4= 1+ \delta x_4$ 
and $x_6=1+\delta x_6$, and working to first order in all deviations, 
\eqref{dmap} yields 
\begin{equation}\label{susysmallm}
\delta m_F' = -\delta m_F ~~\delta m_B'^2 = - \delta m_B^2
\end{equation}
and 
\begin{equation}\label{susysmallo}
\delta x_6' = -\delta x_6 ,~~\delta x_4' =- \delta x_4,~~\delta b_4'=-\delta b_4. 
\end{equation}
It follows from \eqref{susysmallm} that under Giveon-Kutasov duality 
\begin{equation} \label{optranss}
\bar\phi\phi \rightarrow - \bar\phi\phi, ~~~\bar\psi\psi \rightarrow -\bar\psi\psi.   
\end{equation}
\eqref{susysmallo} independently imply that under the  duality  
\begin{equation} \label{optranso}
(\bar\phi\phi)^3\rightarrow -(\bar\phi\phi)^3,~(\bar\phi\phi)^2 \rightarrow (\bar\phi\phi)^2,~ ~   
(\bar\psi \psi) (\bar\phi\phi) \rightarrow (\bar\psi \psi) (\bar\phi\phi).   
\end{equation}
The consistency of \eqref{optranso} with \eqref{optranss} (using large $N$ 
factorization) may be regarded as a mild check of 
the duality map \eqref{dmap}.

The class of theories \eqref{generalaction} also includes a two parameter 
set of ${\cal N}=1$ theories. These theories may be obtained by 
plugging the most general renormalizable superpotential 
$W=-\frac{w}{4k} (\bar \phi \phi)^2-\mu(\bar \phi \phi)$ into the 
general construction of ${\cal N}=1$ theories presented in  equation 
(F.30) of \cite{Jain:2012qi}. The Lagrangian takes the form% 
\footnote{We thank O. Aharony for helping us to correct a mistake  
in an earlier analysis of the ${\cal N}=1$ manifold of theories.}
\eqref{generalaction} with 
$$m_F=\mu, m_B^2=\mu^2, b_4= \mu w, x_4=\frac{1+w}{2}, 
x_6=w^2, y_4'=w, y_4''=w-1.$$ 
Plugging these special values into the general formulas of our paper, 
we have verified that $c_{B,0}=c_{F,0}$, i.e. the zero temperature 
fermionic and bosonic pole masses are equal.  Moreover this two dimensional ${\cal N}=1$ 
submanifold is mapped to itself by the duality transformation \eqref{dmap} 
with the parameter transformations  
$$w' =\frac{3-w}{1+w}, ~~~\mu'= -\frac{2 \mu}{1+w}.$$
We regard the equality of pole masses and the invariance of the 
${\cal N}=1$ submanifold of theories as a nontrivial check of the formulae 
for pole masses and the duality map \eqref{dmap} 
presented in this paper.

The Lagrangian \eqref{generalaction} has three dimensionful parameters; 
$b_4$, $m_F$ and $m_B^2$. It turns out to be possible to scale 
certain combinations of these parameters to infinity while holding others 
fixed, in such a way that the resultant theory is nontrivial and well defined. 
We now describe three such scaling limits. 

In the `fermionic' scaling limit we scale 
$m_F, m_B$ and $b_4$ to infinity as 
\begin{equation} \begin{split} \label{fermionic}
&m_F \to \infty \\
&m_B^2=a_1 m_F^2 + a_2 m_F +a_3\\
&b_4=g_1 m_F +g_2 
\end{split}
\end{equation}
where $x_4$, $x_6$, $a_1, a_2, a_3, g_1, g_2$ are held fixed. We design our 
scaling limit to ensure that the fermionic pole mass $c_{F,0}$ stays fixed 
at a particular value ${\tilde c}_{F,0}$ while $c_{B,0}$ is taken to infinity. 
It turns out that this requirement allows us to solve for $a_1$, $a_2$ and 
$a_3$ as a function of the other parameters. In other words our 
fermionic scaling limit is characterized by one physical parameter  
${\tilde c}_{F,0}$ and four spurious or unphysical parameters, $g_1, g_2, 
x_4, x_6$. At leading order in $\frac{1}{m_F}$, it turns out that the 
thermal free energy of our system is insensitive to the spurious parameters. 
Moreover it exactly matches the free energy of 
single massive fermion coupled to a Chern Simons gauge field
\beal{
S  &= \int d^3 x  \biggl[i \varepsilon^{\mu\nu\rho} {k \over 4 \pi}
\Tr( A_\mu\partial_\nu A_\rho -{2 i\over3}  A_\mu A_\nu A_\rho)
 + \bar\psi \gamma^\mu \wt D_\mu \psi +m_F^{\text{reg}} \bar\psi \psi 
 \biggl]
\label{regular} 
}
with\footnote{\eqref{mr} simply expresses the relationship of the pole mass and the 
bare mass of the theory \eqref{regular}.} 
\begin{equation}\label{mr}
m_F^{\text{reg}}=\left( {\rm \sgn}(m_F^{\text{reg}})-\lambda \right) |c_{F,0}|.
\end{equation}
 Motivated by this result we 
conjecture that  \eqref{generalaction} reduces to \eqref{regular} in
the fermionic scaling limit. 

A second natural scaling limit, which we call the bosonic limit, 
is one in  which  $m_F$, $m_B$ and $b_4$ are also scaled to infinity 
as \eqref{fermionic}, but with parameters chosen in such a manner that  
$c_{F,0}$ is taken to $\infty$ with $c_{B,0}$ and all dimensionless 
couplings ($x_4$ and $x_6$) held fixed. As in the case of the fermionic 
limit, this condition may be used to determine $a_1, a_2$ and $a_3$ as
functions of $g_1$ and $g_2$ (see \eqref{bmbs}). Once again we have checked 
by direct computation that the free energy of our system in this limit 
is independent of the four spurious parameters $g_1, g_2, x_4, x_6$
and in fact agrees exactly with the free energy of the critically 
coupled boson theory 
\beal{
S  &= \int d^3 x  \biggl[i \varepsilon^{\mu\nu\rho} {k \over 4 \pi}
\Tr( A_\mu\partial_\nu A_\rho -{2 i\over3}  A_\mu A_\nu A_\rho)
+ \wt D_\mu \bar \phi \wt D^\mu\phi+\sigma({\bar\phi} \phi+\frac{m^{\text{cri}}_{B}}{4\pi} ) 
 \biggl]
\label{criticalbos}
}
with 
$$|c_{B,0}|=m^{\text{cri}}_B.$$
This result motivates us to conjecture that \eqref{generalaction} reduces 
 to \eqref{criticalbos} in the bosonic scaling limit. It is of course 
natural that our system reduces to a purely bosonic theory when the fermion
pole mass is scaled to infinity. The reader might, however, wonder why 
the bosonic theory so obtained is the critical theory rather than the 
`regular' bosonic theory. We believe that the reason is simply because 
the theory of a `regular' boson with a fixed mass is finely tuned. 
The original bosonic theory we started with has a relevant operator, 
namely $\phi^4$,  whose coefficient is finely tuned to be unnaturally small 
under the 't Hooft limit. 
Integrating out a very massive fermion generically renormalizes the 
coefficient of this operator by a term of order $c_{F,0}$, which is 
very large in the decoupling limit under consideration. However 
the `regular' boson theory perturbed by a $\phi^4$ term with a large 
coefficient is precisely the critical boson, explaining why the 
bosonic scaling limit yields the critical boson theory.

A third `critical' scaling limit we study in this paper is one in which we 
scale $b_4 \to \infty$ with $c_{F,0}$, $c_{B,0}$, $x_4$, $x_6$ held fixed. 
In terms 
of the parameters in \eqref{generalaction} this is achieved by scaling 
$m_B$ to infinity with 
\begin{equation}\label{csl}
b_4=\frac{1}{2 m_B^{\text{cri}} \lambda} m_B^2 + g_2
\end{equation}
where $m_B^{\text{cri}}$ is  held fixed and $g_2$ given in \eqref{css}. 
$m_F$, $x_4$ and $x_6$ are also held fixed
in the critical scaling limit. 
We have verified by explicit computation that $x_4$ and $x_6$ are spurious 
in this limit (in the sense that the finite temperature free energy 
is independent of these parameters). We have also verified that the 
free energy is identical to that of the system 
\beal{
S  &= \int d^3 x  \biggl[i \varepsilon^{\mu\nu\rho} {k \over 4 \pi}
\Tr( A_\mu\partial_\nu A_\rho -{2 i\over3}  A_\mu A_\nu A_\rho)
+ \wt D_\mu \bar \phi \wt D^\mu\phi + \bar\psi \gamma^\mu \wt D_\mu \psi \nn
&
+\sigma({\bar\phi} \phi+\frac{m^{\text{cri}}_{B}}{4\pi} )+{m_F^{\text{reg}}} \bar\psi \psi \biggl].
%&
%+ \frac{4 \pi x_4}{k} (\bar\psi \psi) (\bar\phi\phi)
%+ \frac{2 \pi x_4'}{k} (\bar\psi\phi)( \bar\phi \psi)
%+ \frac{2 \pi x_4''}{k} \left((\bar\psi \phi)( \bar \psi \phi ) 
%+(\bar \phi \psi)( \bar \phi \psi )\right)
%+ \frac{4 \pi^2 x_6}{k^2} (\bar\phi\phi)^3  \biggl].
\label{crit}
}
These results lead us to conjecture that \eqref{generalaction} 
in the scaling limit \eqref{csl} reduces precisely to the 
theory \eqref{crit}. This conjecture is very natural; as we have explained 
above we expect a regular boson to turn into a critical boson in the 
limit that $b_4$ is taken to infinity; once this happens 
 the coefficients $x_4$ and $x_6$ are irrelevant (they multiply operators of 
dimension 4 and 6 respectively) and should make no contribution to any 
physical quantity.

We will now explain how the three different scaling limits identified 
above transform under the duality map \eqref{dmap}. It is not difficult 
to verify that \eqref{dmap} turns the fermionic scaling limit into the 
bosonic scaling limit with $m_{B}^{\text{cri}}$ and $m_F^{\text{reg}}$ related by  
\begin{equation}\label{critdual}
{m_{B}^{\text{cri}}} = -\frac{1}{\lambda -\rm{sgn}(\lm)} m_{F}^{\text{reg}}
\end{equation}
(here $\lambda$ is the 't Hooft coupling of the fermionic scaling limit). 
In a similar manner, \eqref{dmap} turns the bosonic scaling limit into the 
fermionic scaling limit with 
$${m}_{F}^{\text{reg}}=-\lambda {m}_{B}^{\text{cri}}$$
(here $\lambda$ is the 't Hooft coupling of the bosonic scaling limit). 
These relations between ${m_{B}^{\text{cri}}}$ and ${m}_{F}^{\text{reg}}$
were already obtained in \cite{Aharony:2012ns}.
Finally the critical scaling limit with parameter $m_B^{\text{cri}}, m_F^{\text{reg}}$  and 
't  Hooft coupling $\lambda$ is mapped, by \eqref{dmap}, into another critical 
scaling limit with parameters 
${\tilde m}_{B}^{\text{cri}}$ and ${\tilde m}_{F}^{\text{reg}}$ with 
\begin{equation}\label{critdualt} \begin{split}
{\tilde m}_{B}^{\text{cri}} &= -\frac{1}{\lambda -\rm{sgn}(\lm)} m_{F}^{\text{reg}}\\
{\tilde m}_{F}^{\text{reg}}&=-\lambda {m}_{B}^{\text{cri}}.\\
\end{split}
\end{equation}

Combining these results with our conjectures for the nature of the decoupling 
limits it follows that the quantum field theories defined by the Lagrangians 
\eqref{regular} and \eqref{criticalbos} 
are level rank dual to each other when their masses are related by \eqref{critdual}.
This is simply the `bosonization' duality conjectured between the 
regular fermion and critical boson theory in
\cite{Aharony:2012nh} (see also \cite{Giombi:2011kc, Maldacena:2011jn,
Maldacena:2012sf, Aharony:2012ns,Jain:2013py,GurAri:2012is,Takimi:2013zca})
It also follows that the quantum theory defined by the Lagrangian \eqref{crit} is self dual under level 
rank duality, once mass parameters are related by \eqref{critdualt}; this 
is a new result. Note that the bosonization duality conjectured in 
\cite{Aharony:2012nh} follows very simply from the self duality of 
\eqref{crit} by decoupling one of the two fields.

As we have emphasized in this introduction, the existence of {\it a} duality 
map for the (renormalizable subset of) the theories \eqref{generalaction} 
follows immediately from the Giveon Kutasov duality of the ${\cal N}=2$ 
superconformal theory even at large but finite $N$. In this paper we have determined 
the precise form of the duality map, and demonstrated that it reduces, in 
an appropriate scaling limit, to the bosonization duality of 
\cite{Aharony:2012nh} (see also 
\cite{Giombi:2011kc, Maldacena:2011jn,
Maldacena:2012sf, Aharony:2012ns,Jain:2013py,GurAri:2012is,Takimi:2013zca}).
While the precise duality map will of course depend on $N$, it should 
be well approximated by \eqref{dmap} at large but finite $N$. 
Our results thus strongly suggest that the bosonization duality of 
\cite{Aharony:2012nh} is correct even at finite $N$ at least when 
$N$ is large  (recall that all calculational evidence 
for this duality has so far been obtained only in the infinite $N$ limit). 

In summary, the duality map \eqref{dmap} on the general class of theories
\eqref{generalaction} is extremely rich. It reduces to the statement of 
Giveon Kutasov duality for ${\cal N}=2$ theories with this field content, 
suggests a new duality between theories of the form \eqref{crit}, 
and reduces to the bosonization duality between Chern Simons coupled 
minimal fermions and critical bosons in the appropriate decoupling limit.

\section{Exact Results for the thermal free energy}

It was argued in \cite{Aharony:2012ns,Jain:2013py} (following the earlier work \cite{Giombi:2011kc,Jain:2012qi,Yokoyama:2012fa}) 
that the finite temperature partition function of a large $N$ Chern Simons 
theory coupled to fundamental matter, on a 2 sphere of volume $V_2$,  
is given by (see eq (1.3) in  \cite{Jain:2013py})
\begin{equation}\label{cseq}
Z=\langle \exp \left[ -T^2 V_2 v[U] \right] \rangle_{CS}
\end{equation} 
where the RHS is an expectation value in pure Chern 
Simons theory on $S^2 \times S^1$, $T$ is the temperature and $U$ the holonomy 
around $S^1$. $v[U]$ is a gauge invariant function of the holonomy; and 
may be thought of as the free energy of the field theory on $R^2$ 
as a function of the holonomy. It turns out to be possible to explicitly compute $v[U]$ in any given matter Chern Simons theory by summing 
a class of planar free energy graphs.

In general the functional $v[U]$ depends on $N$, $k$ and the parameters of 
the matter Chern Simons theory. Consider two different theories whose 
$v[U]$ functions are given by $v^1_{N, k}[\rho]$ and $v^2_{N', k'}[\rho]$ 
respectively (here $\rho$ is the eigenvalue density function of the holonomy 
matrix $U$). Using the level rank duality of pure Chern Simons theory, the 
authors of \cite{Jain:2013py} were able to demonstrate that if  
\begin{equation} \label{de}
v^1_{N, k}[\rho]=v^2_{|k|-N,-k}[{\rho'}]
\end{equation}
where 
\beal{
\lambda' \rho'(\alpha) = - {\sgn(\lambda) \over 2\pi} + \lambda \rho(\alpha+\pi)
\label{rhotransform}
} 
then the partition function of the theory $1$ at level $k$ and rank $N$ 
is identical to the partition function of the theory $2$ at level $-k$ and 
rank $|k|-N$. In other words the two theories are related by level rank 
duality. In this section we will explicitly compute $v[\rho]$ for the 
theories \eqref{generalaction}. We demonstrate that our results 
\eqref{dmap} obey \eqref{de} when the two different theories ($1$ and $2$) 
are related by the duality map \eqref{dmap}.

By performing the path integral 
over the gauge fields, it was shown in  \cite{Jain:2013py} that 
\eqref{cseq} reduces, in the large $N$ limit, to the saddle point 
solution to the capped matrix model 
(see equations (1.6) and (1.7) in \cite{Jain:2013py})
\begin{equation} \label{ftpf}
Z=\int DU {\rm exp}\left[-V_2 T^2 v[U] \right].
\end{equation}
The integral 
measure $DU$ is capped in a sense we now explain. 
$v[U]$ is most conveniently regarded as a functional of the 
eigenvalue density of the unitary matrix $U$ (see \cite{Jain:2013py} for 
definitions and more details). The integral over unitary matrices in 
\eqref{ftpf} may be thought of as an integral over eigenvalue density functions
that obey the restriction $\rho(\alpha) \leq \frac{1}{2 \pi |\lambda|}$. 
As an eigenvalue distribution is intrinsically positive, we also have 
the inequality $\rho(\alpha) \geq 0$. The result \eqref{ftpf} is 
accurate in the large $N$ limit when $V_2T^2$ is of order $N$ and when 
all dimensionful parameters in the Lagrangian, 
in units of the inverse radius of the sphere,  scale like $N^{\frac{\Delta}{2}}$, 
($\Delta$ is the dimension of the parameter)  

Given the explicit form of $v[\rho]$ it turns out not to be difficult to 
solve the large $N$ saddle point equations that follow from the capped 
matrix integral 
\eqref{ftpf} \cite{Jain:2013py, Takimi:2013zca}. The solution to this 
matrix model determines 
the saddle point value of the eigenvalue density function $\rho(\alpha)$. 
The saddle point value of $\rho(\alpha)$ always tends to a universal 
form in the high temperature (or large volume) limit, 
$\frac{V T^2}{N} \gg 1$,  given by \cite{Jain:2013py} 
\begin{equation}\label{univ} \begin{split}
\rho_{univ}(\alpha) &= \frac{1}{2 \pi |\lambda|}  ~~~( |\alpha| \leq \pi |\lambda|)\\
\rho_{univ}(\alpha) &= 0 ~~~( |\alpha| > \pi |\lambda|).\\
\end{split}
\end{equation}
In this limit the theory effectively lives on $R^2$ and the thermal free energy 
is simply given by $V T^2 v[\rho_{univ}]$. 

In this section we present explicit results for the function $v[\rho]$ for 
the theory \eqref{generalaction}. Our result for $v[\rho]$  depends on 
five field theory parameters divided by appropriate powers of the temperature. 
These parameters are 
\begin{equation}\label{resca}
{\hat m}_B^2=\beta^2 m_B^2, ~~~{\hat m}_F=\beta  m_F , ~~~
{\hat b}_4= \beta b_4,~~~ x_4,~~~ x_6
\end{equation} 
($\beta$ is the inverse temperature). 
The results also depend on the two independent non dimensionalized 
chemical potentials 
$\nu_F= \beta\mu_F$ and $\nu_B=\beta \mu_B$. 
Here $\mu_B$ and $\mu_F$ are the chemical potentials 
for the symmetries under which $\phi \to e^{i\alpha} \phi$ and 
$\psi \to  e^{i \alpha} \psi$ respectively.% 
\footnote{The Lagrangian \eqref{generalaction} is invariant under separate rephasings 
of the bosonic and fermionic fields provided $y_4''=0$. It turns out that the 
free energy of our system is independent of $y_4''$ at leading order in 
the large $N$ limit studied in this paper. For this reason it appears
to be consistent to turn on independent chemical potentials for 
bosons and fermions, for any field theory of the form \eqref{generalaction}, 
in the large $N$ limit studied in this paper.}

The methods we have employed to obtain our results are identical to those 
employed in \cite{Giombi:2011kc,Jain:2012qi,Aharony:2012ns,Jain:2013py}, see appendix \ref{eea} for details of the computation. The procedure that was   
used by \cite{Giombi:2011kc,Jain:2012qi,Aharony:2012ns,Jain:2013py} to obtain the thermal free energy 
also evaluates the thermal masses of the bosonic and fermionic fields. Below 
present explicit results for these thermal masses in addition to the 
free energy functional $v[\rho]$. In the zero 
temperature limit these masses reduce to the physical scattering pole 
masses of the bosonic and fermionic fields respectively.

\subsection{Results for pole masses and thermal free energy}

In this section we describe our results for the pole masses and finite 
temperature free energies of the action \eqref{generalaction}.% 

In the lightcone gauge of \cite{Giombi:2011kc}, which we use in this paper, the 
exact propagators for the bosonic and fermionic fields in our problem, 
take the form 
\begin{equation} \label{prop}
\begin{split}
\langle {\bar \phi}_j(-q) \phi^i(p) \rangle &= \frac{(2 \pi)^3 \delta^i_j \delta^3(-p+q)}
{\wt p^2+ c_B^2 T^2} \\
\aver{\bar\psi_j(-q) \psi^i(p)} &= {- \delta^i_j  (2 \pi)^3 \delta^3(-p+q)\over i\gamma^\mu \wt p_\mu + \wt \Sigma_F(p)}.
\end{split}
\end{equation}
We work in the finite temperature Euclidean theory in which the direction 
$x^3$ is compactified on a circle of circumference $\beta$. In our 
expression for the bosonic propagator we have used the symbol $\wt p_\mu$ 
whose definition is    
\be\wt p_\mu = p_\mu+i \delta_{\mu,3}\mu_{B}.\ee
In a similar fashion the symbol $\wt p_\mu$ that appears in the fermionic 
propagator is defined by 
\be\wt p_\mu = p_\mu+i \delta_{\mu,3}\mu_{F}.\ee
 
Upon solving the gap equation that determines these exact propagators in the 
large $N$ limit we find that $\wt \Sigma_F(p)$ is given by 
\beal{ 
{\wt \Sigma_F}(p)=& f(\hat p_s) p_s + i g(\hat p_s ) p_+ \gamma^+ \nn
f (\hat p_s) =&{\lambda \over \hat p_s} \int d\alpha\rho(\alpha)   \biggl\{ \log(2 \cosh \frac{\sqrt{\hat p_s^2 +c_F^2} +i\alpha + \nu_F}{2})+ \log(2 \cosh \frac{\sqrt{\hat p_s^2 +c_F^2} - i\alpha - \nu_F}{2}) \nn
&- x_4  \( \log(2 \cosh \frac{c_B +i\alpha + \nu_B}{2})+ \log(2 \cosh \frac{c_B - i\alpha - \nu_B}{2})  \) \biggl\} + {\hat m_F \over \hat p_s} \nn
g(\hat p_s) =&\frac{c_F^2}{\hat p_s^2}- f(\hat p_s)^2 
\label{sigmafg}
}
where $\alpha$ represents holonomy and $\rho(\alpha)$ is holonomy distribution, and $\hat p_\mu$ is the non dimensionalized momentum
\be
\hat p_\mu = {p_\mu \over T}.
\ee
As in \cite{Giombi:2011kc} we use the symbol $p_s$ to denote $\sqrt{p_1^2+ p_2^2}$ 
(see \cite{Giombi:2011kc} for more details).

The constants $c_B$ and $c_F$ that appear in \eqref{prop} 
have the interpretation of the thermal mass (in units of the temperature) 
of the bosonic and fermionic fields respectively. These quantities 
are determined as solutions to the equations
\beal{
c_F^2=& \left\{2 {\lambda} (\cC - x_4 \cS ) +\hat m_F \right\}^2 
\label{girfreesquare}\\
c_B^2 
=& \lambda^2 (1 + 3 x_6) \cS^2 -4\lambda \hat b_4 \cS + 4 x_4 \(  \lambda^2\cC^2 - 2 x_4 \lambda^2 \cS \cC +\hat m_F  \lambda \cC \) +\hat m_B^2
\label{bosfree}
}
where hatted variables were defined in \eqref{resca} and%
\footnote{The gap equations \eqref{girfreesquare} and \eqref{bosfree} 
both involve only functions of $|c_F|$ rather than $|c_F|$ itself. 
It follows that if $(c_F, c_B)$ is a solution to the gap equations 
then $(-c_F, c_B)$ is also a solution to the same equation. However these 
two related solutions actually specify the same physical solution, as 
the propagator and the free energy are functions of $|c_F|$ 
rather than $c_F$ itself. This fact may be emphasized by writing 
\eqref{girfreesquare} 
may be rewritten as an equation for $|c_F|$ as 
$$|c_F| = \left| 2 {\lambda} (\cC - x_4 \cS ) +\hat m_F \right| $$
(note that the RHS of this equation is a function only of $|c_F|$). 
Alternately we could rewrite \eqref{girfreesquare} as 
$$c_F =  2 {\lambda} (\cC - x_4 \cS ) +\hat m_F $$
(where we have chosen a particular branch of the square root, quite 
at random; which branch we choose has no physical significance).  
Both these forms of the equation capture all of its physical content. 
The last form makes it clear that the solution is in fact an analytic 
function of field theory parameters as $c_F$ crosses zero.}
\beal{
\cC =& \half \int d\alpha\rho(\alpha)   \( \log(2 \cosh \frac{|c_F| +i\alpha + \nu_F}{2})+ \log(2 \cosh \frac{|c_F| - i\alpha - \nu_F}{2}) \)
\label{cc}\\
\cS =& \half \int d\alpha\rho(\alpha)   \( \log(2 \sinh \frac{|c_B| +i\alpha + \nu_B}{2})+ \log(2 \sinh \frac{|c_B| - i\alpha - \nu_B}{2}) \).
\label{ss}
}

The gap equations presented above simplify in the zero temperature limit. 
This limit is taken by setting 
$$c_B= \beta c_{B,0}, ~~~c_F= \beta c_{F,0}, ~~~\nu_F=\beta \mu_F, ~~~
\nu_B=\beta \mu_B$$
in the equations above and taking $\beta $ to infinity holding 
$c_{B,0}$, $c_{F,0}$, $\mu_F$ and $\mu_B$ fixed.% 
\footnote{Note that $c_{B,0}$ 
and $c_{F, 0}$ are simply the dimensionful zero temperature 
pole masses of the boson and fermion 
respectively, while $\mu_F$ and $\mu_B$ are the dimensionful fermionic 
and bosonic chemical potentials.} 
In this limit 
\begin{equation} \label{ztr}
\beta {\cal S} \rightarrow \frac{ {\rm max}(|c_{B, 0}|, |\mu_B|) }{2}, 
~~~\beta {\cal C} \rightarrow \frac{ {\rm max}(|c_{F, 0}|, |\mu_F|) }{2}.
\end{equation}
Using these replacement rules it is easily verified that the gap equations 
reduce in this limit to \eqref{fpm} and \eqref{ztm} reported in the 
introduction in the case that $|c_F|>|\mu_F|$ and $|c_B|>|\mu_B|$; the 
equations in the other cases may be worked out as easily.% 
\footnote{It is possible to explicitly solve \eqref{fpm} and 
\eqref{ztm} to obtain the zero temperature pole masses as a function 
of the parameters of the Lagrangian. This process requires us to solve 
a quadratic equation; in general  we find two inequivalent branches of solutions. 
Although we have not analysed this question, a similar situation presumably 
persists at finite temperature. The thermodynamically preferred 
branch is the one that has the lower free energy. It would be interesting 
to analyse the space of solutions, and thermodynamical properties, of the 
finite temperature gap equations.}

Returning to the finite temperature computation, the function $v[\rho]$ (which 
determines thermal free energy of our system on a sphere via \eqref{ftpf}, 
) is given by
\beal{
\!\! \!{ 6 \pi v[\rho] \over N} 
=& - |c_B|^3 + 2 (c_B^2 - \hat m_B^2 )\cS   + 3 \int d\alpha \rho(\alpha) \int_{|c_B|} ^\infty dy y \( \log (1 - e^{-y-i\alpha -\nu_B}) +\log (1 - e^{-y+i\alpha +\nu_B}) \) \nn
& + |c_F|^3 - 2 (c_F^2 - \hat m_F^2) \cC
- 3  \int d\alpha \rho(\alpha) \int_{| c_F|} ^\infty dy y \( \log (1 + e^{-y-i\alpha -\nu_F }) + \log (1 + e^{-y + i\alpha + \nu_F }) \) \nn
&+ 2 \lambda \hat b_4 \cS^2 + 2 \lambda \hat m_F \cC^2 - 4\lambda x_4\hat m_F\cS\cC \nn
&+ |\hat c_{B,0}|^3+ |\hat c_{B,0}| \hat m_{B}^2
-|\hat c_{F,0}|^3 - |\hat c_{F,0}| \hat  m_{F}^2-\frac{\lambda }{2} \hat b_4\hat c_{B,0}^2 
-\frac{\lambda}{2}\hat m_F \hat c_{F,0}^2   + x_4 \lambda \hat m_{F} | \hat c_{B,0}| |\hat  c_{F,0}|,
\label{free1}
}
where we have defined 
\be \hat c_{F,0}=\beta c_{F,0},~~~\hat c_{B,0}=\beta c_{B,0}.
\ee
Terms in $v[\rho]$ above that scale like $\beta^3$ (and are independent 
of all chemical potentials and the holonomy) are ambiguous; 
they can be modified by adding 
a cosmological constant counter term to the field theory action. Physically 
this ambiguity affects only the vacuum energy of the theory; 
such an ambiguity is always present in any quantum field theory decoupled
from gravity. We have chosen to fix this ambiguity by a choice of cosmological 
constant counterterms that sets the zero temperature 
vacuum energy of our theory to zero. 

In obtaining the relatively simple expression \eqref{free1} for $v[\rho]$, 
we have made liberal use of the gap equations. In that sense 
\eqref{free1} is an `on-shell' expression for the free energy. It is possible 
to add terms to \eqref{free1} that vanish on-shell (i.e. when the gap 
equations are satisfied) to obtain an off-shell `Landau Ginzburg' 
free energy that has the following interesting property: the 
gap equations may be derived from this off-shell 
free energy by extremizing it w.r.t $c_B$ and $c_F$.% 
\footnote{The existence of such an off-shell free energy follows from 
the derivation of the gap equations presented in  \cite{Jain:2012qi}; 
the gap equations were derived there by varying the free energy w.r.t 
the $\alpha$ and $\Sigma$ variables defined in that paper. }
This off-shell free energy may be regarded as a Landau Ginzburg free energy, 
which is a function of the order parameters $c_B$ and $c_F$.

Algebraically, the off-shell free energy may be obtained as follows. We 
first view the equations \eqref{girfreesquare} and \eqref{bosfree} as 
equations that determine ${\cal S}$ and ${\cal C}$ as functions of 
$c_F$ and $c_B$. Solving these equations yields ${\cal S}={\cal S}(c_F, c_B)$ 
and ${\cal C}={\cal C}(c_B, c_F)$. We then plug these solutions into 
every explicit occurrence of ${\cal S}$ and ${\cal C}$ in \eqref{free1}. 
This procedure may be implemented explicitly, and yields the complicated 
but explicit expression for $v[\rho]$ given in Appendix \ref{freeenergy}. 
The expression for $v[\rho]$ so obtained is clearly identical to 
\eqref{free1} on-shell; moreover it turns out that its variation yields 
the gap equations. We have checked this assertion using the explicit 
expression presented in Appendix \ref{freeenergy}; it is also 
possible to formally verify this claim (without using the explicit 
expressions of Appendix \ref{freeenergy}) as follows. 
The derivative of the off-shell $v[\rho]$ w.r.t $c_B$ and $c_F$ is given by 
\beal{
\!\! \!{ 6 \pi  \over N} { \partial v[\rho] \over \partial |c_B|}
=& 2 \left(c_B^2-m_B^2\right) {\partial \cS \over \partial |c_B|} -2 |c_B| \cS 
-2 {\partial \cC \over \partial |c_B|} \left(c_F^2-m_F^2\right) \nn
&+4 \lambda b_4 \cS {\partial \cS \over \partial |c_B|}  +4 m_F \lambda \cC {\partial \cC \over \partial |c_B|} 
-4 x_4 \lambda m_F ({\partial \cC \over \partial |c_B|} \cS+\cC {\partial \cS \over \partial |c_B|}) \nn
\!\! \!{ 6 \pi  \over N} { \partial v[\rho] \over \partial |c_F|}
=&-2 {\partial \cC \over \partial |c_F|} \left(c_F^2-m_F^2\right)+2 \cC |c_F|   
+2 \left(c_B^2 - m_B^2\right) {\partial \cS \over \partial |c_F|} \nn
& +4\lambda  b_4 \cS {\partial \cS \over \partial |c_F|} +4 m_F \lambda \cC {\partial \cC \over \partial |c_F|} 
-4 x_4 \lambda m_F ({\partial \cC \over \partial |c_F|} \cS+\cC {\partial \cS \over \partial |c_F|}). 
}
We have shown that the right-hand side of both these expressions 
vanish by using \eqref{girfreesquare} and 
\eqref{bosfree} (and their derivative of these equations w.r.t $c_B$ and $c_F$).
In other words the variation of our off-shell $v[\rho]$ w.r.t. $c_B$ and $c_F$ 
vanishes when the gap equations are satisfied, i.e. the gap equations 
follow from the variation of the off-shell action.

\subsection{Invariance of the gap equations under duality}

In this subsection we will demonstrate that, under appropriate conditions, 
the gap equations 
\eqref{girfreesquare} and \eqref{bosfree} are invariant under the duality 
map \eqref{dmap} provided we interchange $c_B$ and $c_F$ as well as  
the bosonic and fermionic chemical potentials $\mu_B$ and $\mu_F$  
and also perform a duality transformation of the holonomy distribution 
\eqref{rhotransform}. In the next subsection we will show that the analogous 
result also holds for the free energy functional $v[\rho]$. As explained 
above, this result guarantees the duality invariance of the $S^2$ thermal 
free energy of the theories \eqref{generalaction} under the duality 
transformation \eqref{dmap}.

The gap equations 
\eqref{girfreesquare} , \eqref{bosfree} and the free energy \eqref{free1} enjoy 
invariance under duality only when the following three conditions are met
\beal{ & \label{co1} |c_B| \geq |\nu_B| \\
& \label{ct1} |c_F| \geq |\nu_F|\\
& \label{cb} \sgn(\lambda) \left\{2 {\lambda} (\cC - x_4 \cS ) +\hat m_F \right\} 
\geq 0.
}
When any of the three conditions listed above are not met, the gap equations and 
free energy presented in this paper do not enjoy invariance under duality. 
We now present a possible interpretation of this fact, following 
\cite{Aharony:2012ns}.

When the condition \eqref{co1} is violated we would, at least naively, expect 
the bosonic field in our theory to condense.% 
\footnote{Condensation of the scalar field could well 
occur even though the finite temperature 
path integral studied in this paper is effectively two dimensional, and 
spontaneous symmetry breaking of a continuous symmetry is forbidden in two 
dimensions. This is because we are studying a gauge theory; the condensation 
phenomenon in this theory is the Higgs phenomenon and does not result in 
a massless Goldstone field.}
In other words when \eqref{co1} is not met it seems likely either that 
the `uncondensed' gap equations presented in this paper have no solution, 
or that if a solution exists it is not the dominant saddle point.% 
\footnote{Indeed 
if such a solution exists it may not even be a legal saddle point of our 
theory - whether dominant or not - as in deriving the saddle point 
equation from the field theory action one appears to have performed 
integrals of the form $\int dx e^{-a x^2}$ with $\text{Re}(a) <0$.}
We suspect that the 
same is true when the the condition \eqref{cb} is not met; in this situation
as well we suspect that the true saddle point of our theory involves a Bose 
condensate.\footnote{This expectation is supported by a slight generalization 
of an argument presented in \cite{Aharony:2012nh, Aharony:2012ns}. Let us 
work at zero 
temperature and chemical potential. In this situation the RHS of \eqref{cb} 
vanishes when the physical mass of the fermionic field vanishes. It follows that the physical fermion mass flips sign when the RHS in \eqref{cb} flips sign. 
The effective level of the low energy Chern Simons theory, obtained by
integrating out the massive fermionic field changes by one unit as the 
fermion flips sign. This suggests that the rank of the dual low energy 
gauge group is reduced by one unit under level rank duality, a phenomenon 
that occurs due to the Higgs mechanism when a scalar condenses. We thank 
O. Aharony for explaining this to us. }

A very similar phenomenon appears to occur in the 
critically coupled scalar field coupled to a Chern Simons gauge field, 
as we have investigated in some detail in Appendix \ref{bosecondensation} 
(see \cite{Aharony:2012ns} for a related discussion).% 
\footnote{We have not carefully investigated the conditions under which 
the saddle point equations presented in this paper have solutions. We 
strongly suspect that solutions exist when the inequalities \eqref{co1}, 
\eqref{ct1} and \eqref{cb} are all met. We also suspect that solutions 
do not always exist when some of these conditions are violated. 
We leave an investigation of this issue to future work.}

In summary, we believe that when either of the conditions \eqref{co1} or 
\eqref{cb} is violated, the expressions for thermal masses and free energies
presented in this paper do not apply.\footnote{It seems possible that the 
correct expressions for the free energy and thermal masses are simply obtained
by analytically continuing the `legal' expressions presented in this paper 
(i.e. the expressions when \eqref{co1} and \eqref{cb} do apply) to values 
of parameters where they are apparently violated. See Appendix \ref{bosecondensation} 
for a discussion of this in a simpler context. } What of the condition 
\eqref{ct1}? This is simply the duality map of the condition \eqref{co1} 
(it is possible to verify that the condition \eqref{cb} is invariant under 
duality). In other words, the gap equations and thermal free 
energy presented in this paper are correct if and only if \eqref{co1} and
\eqref{cb} apply both in the original theory and in the dual theory. 
This is what we would expect if it is indeed true that the correct saddle 
point is different from the one presented in this paper if either 
\eqref{co1} or \eqref{cb} are violated. 

Provided that \eqref{co1} and \eqref{cb} are obeyed, the expressions 
$\cS$ and $\cC$ in \eqref{ss} and \eqref{cc} transform under duality as  
\beal{
\lambda' \cS' = -{ \sgn(\lambda) \over 2 } |c_F| +\lambda \cC, \quad 
\lambda' \cC' = -{ \sgn(\lambda) \over 2 } |c_B| +\lambda \cS 
\label{sscctransform}
}
(here unprimed quantities refer to the original theory and primed quantities 
to the dual theory). Using these relations and \eqref{cb}, the dual equation
\be
(c_F')^2= \left\{2 {\lambda'} (\cC' - x_4' \cS' ) +\hat m_F' \right\}^2
\label{pe}
\ee
may easily be shown to reduce to \eqref{girfreesquare}, as we now explain. 
We first rewrite \eqref{pe} as
\begin{equation}\label{pee}
|c_F'|= {\rm sgn}(X')\left\{2 {\lambda'} (\cC' - x_4' \cS' ) +\hat m_F' \right\}
\end{equation} 
where 
$$X'=(\cC ' - x_4' \cS' ) +\hat m_F'.$$
Using \eqref{sscctransform} (which, recall is valid provided that 
\eqref{co1} and \eqref{ct1} apply), we may rewrite \eqref{pee} as 
\begin{equation} 
|c_B|= -{\rm sgn}(\lambda) {\rm sgn}(X')|c_B|
+ \frac{{\rm sgn X'}}{x_4} 
\left[ {\rm sgn}(\lambda) |c_F| - \left( 2 \lambda ( {\cal C}-x_4 {\cal S} ) 
+{\hat m_F} \right) \right]. \label{pet}
\end{equation} 
If the inequality \eqref{cb} applies to the primed theory then \eqref{pet} 
simplifies (recall ${\rm sgn}(\lambda)=-{\rm sgn}(\lambda')$) to 
\begin{equation}\label{peto} 
2 \lambda \left( {\cal C}-x_4 {\cal S} \right) 
+{\hat m_F} ={\rm sgn }(\lambda)|c_F|.
\end{equation}
Clearly the RHS and LHS of \eqref{peto} have the same sign if and 
only  
\begin{equation} \label{condd}
{\rm sgn}(\lambda) {\rm sgn}(X)=1.
\end{equation}
We assume that \eqref{pe} has a solution. This implies that \eqref{peto}
has a solution (as \eqref{peto} is simply a rewriting of \eqref{pe}), and 
so it follows that \eqref{condd} is met. In other words we have used 
the inequality \eqref{cb} (and other conditions) for the primed theory to 
derive the inequality \eqref{cb} for the unprimed theory, justifying the 
claim above that \eqref{cb} is a `selfdual' condition. Squaring \eqref{peto}
we now recover \eqref{girfreesquare}, the unprimed version of \eqref{pe}, as 
we set out to show.

 By plugging 
\eqref{sscctransform} into the primed version of  \eqref{bosfree}
\beal{
c'_B{}^2 
=& \lambda'{}^2 (1 + 3 x'_6) \cS'{}^2 -4\lambda' \hat b'_4 \cS' + 4 x'_4 \(  \lambda'{}^2\cC'^2 - 2 x'_4 \lambda'{}^2 \cS' \cC' +\hat m'_F  \lambda' \cC' \) +\hat m'_B{}^2
}
and eliminating ${\cal C}$ using \eqref{girfreesquare} one also recover 
\eqref{bosfree} (we do not present the details of the straightforward 
algebra here). In other words the gap equations enjoy invariance under 
duality under the appropriate conditions. The duality invariance of the gap equations at zero temperature  \eqref{ztm},\eqref{fpm} follow
by taking zero temperature limit.

\subsection{Invariance of the free energy under duality}

In this subsection we will illustrate the free energy \eqref{free1} is also invariant 
under the duality transformation \eqref{dmap}, provided 
we also simultaneously interchange $c_B$ and $c_F$ and $\nu_B$ and $\nu_F$ with \eqref{rhotransform}. 

We find it convenient to study the duality transformation of $v[\rho]$ 
after dividing it by $k$; in other words we study the transformation 
of 
\beal{
\!\! \! {6 \pi   v[\rho] \over k}  
=& - \lambda |c_B|^3 + 2 (c_B^2 - \hat m_B^2 )\lambda\cS   + 3 \int d\alpha \lambda\rho(\alpha) \int_{|c_B|} ^\infty dy y \( \log (1 - e^{-y-i\alpha -\nu_B}) +\log (1 - e^{-y+i\alpha +\nu_B}) \) \nn
& + \lambda |c_F|^3 - 2 (c_F^2 - \hat m_F^2) \lambda\cC
- 3  \int d\alpha \lambda\rho(\alpha) \int_{ |c_F|} ^\infty dy y \( \log (1 + e^{-y-i\alpha -\nu_F }) + \log (1 + e^{-y + i\alpha + \nu_F }) \) \nn
&+ 2 \hat b_4 (\lambda\cS)^2 + 2 \hat m_F (\lambda\cC)^2 - 4 x_4\hat m_F\lambda^2 \cS\cC \nn
&+\lambda| \hat c_{B,0}|^3+\lambda|\hat c_{B,0}| \hat m_{B}^2
-\lambda|\hat c_{F,0}|^3 - \lambda|\hat c_{F,0}| \hat  m_{F}^2-\frac{1}{2} \hat b_4(\lambda \hat c_{B,0})^2 
-\frac{1}{2}\hat m_F (\lambda\hat c_{F,0})^2   + x_4 \lambda^2 \hat m_{F}  |\hat c_{B,0}|| \hat  c_{F,0}|.
\label{free2}
}
We will demonstrate that the RHS of this expression maps to  minus itself 
under duality. This ensures $v[\rho]$ maps to itself under duality (recall 
 since $k$ goes to $-k$ under duality).  
 
In order to demonstrate the invariance of the RHS under duality, upto a minus 
sign, we find it convenient to decompose the RHS of \eqref{free2} into 
four parts; RHS=$f_1+f_2+f_3 +f_4$ where 
\beal{
f_1 =& - \lambda |c_B|^3 + 2 c_B^2 \lambda\cS   + \lambda |c_F|^3 - 2 c_F^2  \lambda\cC\\
f_2 =&   3 \int d\alpha \lambda\rho(\alpha) \int_{|c_B|} ^\infty dy y \( \log (1 - e^{-y-i\alpha -\nu_B}) +\log (1 - e^{-y+i\alpha +\nu_B}) \) \nn 
&- 3  \int d\alpha \lambda\rho(\alpha) \int_{ |c_F|} ^\infty dy y \( \log (1 + e^{-y-i\alpha -\nu_F }) + \log (1 + e^{-y + i\alpha + \nu_F }) \) \\
f_3=& -2 \hat m_B^2 \lambda\cS + 2 \hat m_F^2 \lambda\cC + 2 \hat b_4 (\lambda\cS)^2 + 2 \hat m_F (\lambda\cC)^2 - 4 x_4\hat m_F\lambda^2 \cS\cC \\
f_4=&\lambda |\hat c_{B,0}|^3+\lambda|\hat c_{B,0}| \hat m_{B}^2
-\lambda|\hat c_{F,0}|^3 - \lambda|\hat c_{F,0}| \hat  m_{F}^2-\frac{1}{2} \hat b_4(\lambda \hat c_{B,0})^2 
-\frac{1}{2}\hat m_F (\lambda\hat c_{F,0})^2   + x_4 \lambda^2 \hat m_{F}  |\hat c_{B,0}|| \hat  c_{F,0}|.
}
It is not difficult to demonstrate that $f_1, f_2$ map to  $-f_1,-f_2$ 
respectively under duality (we use \eqref{sscctransform}, 
\eqref{rhotransform} and also the transformation of $c_B$ and $c_F$, 
$\mu_B$ and $\mu_F$). 
The algebra involved in the demonstration of the duality invariance of 
$f_3$ and $f_4$ is more lengthy, and we used Mathematica. We find that% 
\footnote{This right-hand value is clearly invariant under the duality map 
\eqref{dmap}. }
\beal{
f_3+ f_3' =& -(f_4 + f_4' )
= \frac{3 \hat m_F \left(-4 \hat b_4 \hat m_F x_4+4 \hat m_B^2 x_4^2+\hat m_F^2 (-1+x_6)\right)}{8 x_4^3}. 
}
In other words $f_3 + f_4$ maps to $-(f_3+f_4)$ under the duality 
transformation.% 
\footnote{
$f_4$ corresponds to zero point energy subtraction so that $v[\rho]=0$ at 
$T=0$.  }
By collecting these results it follows that $(v[\rho])'=v[\rho]$.
This completes our demonstration of invariance of the thermal free energy under the duality transformation. 

\section{Scaling limits} 

As we have explained in the introduction, the general Lagrangian 
\eqref{generalaction} simplifies in appropriate scaling limits, in 
which the dimensionful parameters $m_F$, $m_B$ and $b_4$  are sent to 
infinity in appropriate ratios. In this section we study some of these 
scaling limits and their interplay under duality transformations. 

\subsection{Fermionic scaling limit}
\label{fermionlimitI}

In this subsection we describe a set of scaling limits under which 
we conjecture that the general renormalizable system \eqref{generalaction} 
reduces to \eqref{regular}, a theory of fermions minimally coupled to the 
Chern Simons gauge field. As evidence for our conjecture we demonstrate that 
the gap equations \eqref{girfreesquare}, \eqref{bosfree} and the free energy 
\eqref{free1} reduce to the corresponding objects for  the regular 
fermion theory \eqref{regular} in the limit presented in this subsection. 

As explained in the introduction, we focus on scaling limits in which 
 all dimensionful Lagrangian parameters are taken to infinity at a rate 
governed by their scaling dimensions, in the manner quantitatively described 
by \eqref{fermionic}. Inserting this limit into the gap equations \eqref{fpm}
allows us to solve for $c_{B, 0}$ and $c_{F,0}$ (in terms of the parameters 
$a_1, a_2, a_3, g_1, g_2$  of \eqref{fermionic}) in a power series expansion 
in $m_F$. In particular we find that  $c_{F,0}$ takes the form 
$A m_F^2+ B m_F + C$. As we wish to tune our scaling limit so that $c_{F,0}$
is held fixed at a particular value - say ${\tilde c}_{F,0}$ - in the limit
$m_F \rightarrow \infty$, we must set $A=B=0$ and $C={\tilde c}_{F,0}$. 
These three conditions
allow us to solve for $a_1$, $a_2$ and $a_3$ in terms of $g_1$, $g_2$ and 
${\tilde c}_{F,0}$ (see \eqref{fermionic} for a definition of these 
quantities).  We find 
\be\begin{split}\label{fmbss}
&a_1= \frac{\lambda ^2 (8 g_{1} x_{4}-3 x_{6}-1)+4}{4 \lambda ^2 x_{4}^2},~~
a_2 = \frac{2 g_{2}}{x_{4}}-\frac{m_F^{\text{reg}} \left(\lambda ^2 (4 g_{1} x_{4}-3 x_{6}-1)+4\right)}{2 \lambda ^2 x_{4}^2}\\
&a_3= (m_F^{\text{reg}})^2\(x_4\frac{\lm (\lm-2\sgn(\text X^{\text{reg}}))}{(\sgn(\text X^{\text{reg}})-\lm)^2}+\frac{\frac{4}{\lm^2}-3x_6-1}{4 x_4^3}\)
-\frac{2 g_{2} m_F^{\text{reg}}}{x_{4}}
\end{split}\ee
with $X^{\text{reg}}= m_F^{\text{reg}}$. 
Recall however that we only trust the results of this paper 
when the inequality \eqref{cb}; at zero temperature in the fermionic 
limit \eqref{cb} reduces to the condition ${\rm sgn}(m_F^{\text{reg}}) =
{\rm \sgn}(\lambda)$. We choose to study the limit \eqref{fmbss} with  
\be
\text {\rm sgn} (X^{\text{reg}})= {\rm sgn}(\lambda).
\ee
(we will explain this choice very soon).

 Now in the regular fermion theory \eqref{regular}, the fermionic pole mass 
 is given in terms of the mass parameter in \eqref{regular} by the equation%
 \footnote{As we have explained above the sign of $c_{F,0}$ and the sign 
of $c_F$ are ambiguous. Throughout this paper we follow the convention that 
$c_{F}$ has the same sign as  
$\left\{2 {\lambda} (\cC - x_4 \cS ) +\hat m_F \right\}$ }
 \be
 {\tilde c}_{F,0} = { m_F^{\text{reg}} \over \sgn(m_F^{\text{reg}}) -\lambda}.
 \label{rfse}
 \ee
As we expect \eqref{generalaction} to reduce to \eqref{regular} in 
the scaling limit of this section, it is convenient to make the substitution
\eqref{rfse} into \eqref{fmbss}, so that our scaling limit is parameterized 
by $m_F^{\text{reg}}$ rather than ${\tilde c}_{F,0}$. 

Let us now turn to the finite temperature gap equations of our system. 
These equations may be simplified by noting that ${\cal S}= \frac{|c_B|}{2}
+ {\cal O}(e^{-c_B})$ in the large $c_B$ limit. Making use of this fact we 
find that the finite temperature gap equations simplify, in the scaling 
limit of this section, to 
\beal{\label{fselfs}
&c_{B} = \frac{\hat m_F-\hat m_F^{\text{reg}} }{x_4 \lm}+\frac{b_1}{\hat m_{F}}+\frac{b_2}{\hat m_{F}^2}+{\cal O}(\frac{1}{\hat m_{F}})^3\nn
&\cC=\frac{\sgn(\lambda)~c_{F}-\hat m_F^{\text{reg}}}{2\lm}+\frac{d_1}{\hat m_{F}}+\frac{d_2}{\hat m_{F}^2}+{\cal O}(\frac{1}{\hat m_{F}})^3\nn
&b_{1} =\frac{ 2 d_1}{x_4} ,~~b_{2} = \frac{ 2 d_2}{x_4}\\
&d_1 = \frac{\lambda  x_{4}^3 \left(c_{F}^2-\frac{(\hat m_F^{\text{reg}})^2}{(\sgn(\lambda)-\lm)^2}\right)}{\lambda ^2 (4 g_{1} x_{4}-3 x_{6}-1)+4}\nn
&d_{2} =\frac{\lambda  x_{4}^3 \left(c_{F}^2-\frac{(\hat m_F^{\text{reg}})^2}{(\sgn(\lambda)-\lm)^2}\right) 
\left(\hat m_F^{\text{reg}} \left(\lambda ^2 \left(4 x_{4}^3-3
   x_{6}-1\right)+4\right)-4 g_{2} \lambda ^2 x_{4}\right)}{\left(\lambda ^2 (4 g_{1} x_{4}-3 x_{6}-1)+4\right)^2}\notag
}
The first equation in \eqref{fselfs} asserts that $c_B$ is the same as 
$c_{B,0}$ upto corrections that vanish in the large $m_F$ limit (we will, 
however need to keep track of some of these corrections in order to get 
the correct value of $v[\rho]$, as we will explain below). The leading order 
piece in the second  equation in \eqref{fselfs} may be rewritten as 
\begin{equation}\label{ses}
\cC=\frac{\sgn(\lambda)~|c_{F}|-\hat m_F^{\text{reg}}}{2\lm}
\end{equation}
(after substituting for ${\tilde c}_{F,0}$ in terms of $m_F^{\text{reg}}$). 
However \eqref{ses} is precisely the gap equation for the regular fermionic 
theory \eqref{regular} (see  \eqref{gap}) provided 
\begin{equation}\label{fermineq}
\sgn(\lm) \left (  m_F^{\text{reg}}+2 \lm \cC \right ) \geq 0.
\end{equation}
It is easily verified that \eqref{fermineq} follows as a consequence of 
\eqref{cb}. In other words the finite temperature gap equation of 
\eqref{generalaction} agrees precisely with the same equation in the 
purely fermionic theory \eqref{regular} whenever our results are reliable 
(i.e. whenever \eqref{cb} is valid).

We now turn to the computation of the off-shell free energy of our 
system in the fermionic scaling limit. Recall that, in general, the 
procedure for computing this off-shell free energy is to use the gap equations 
to solve for ${\cal S}$ and ${\cal C}$ in terms of $c_F$ and $c_B$ and 
to substitute these expressions into \eqref{free2}. This procedure is 
particularly easy to implement in the scaling limit of this section. 
The required solutions are given in the first two equations of 
\eqref{fselfs} (recall that ${\cal S}= \frac{|c_B|}{2}$ in the limit under 
study). Plugging these into \eqref{free2} and collecting terms%
\footnote{Note that ${\cal S}$ and ${\cal C}$ in \eqref{free2} are multiplied by 
terms that diverge like $m_F^2$ in the scaling limit considered in this paper. 
For this reason it was very important to keep all subleading terms in the 
first two equations of \eqref{fselfs}, upto order ${\cal O}(1/m_F^2)$. Omitting
these terms would lead to a divergence at order $m_F$ and errors in the 
finite piece.}
we find 
that all divergent pieces in $v[\rho]$ cancel leaving a finite result given 
by 
\beal{\label{fefe}
&v[\rho]\nn
&= {N  \over 6 \pi}  \biggl[ { | c_{F}|^3 } \frac{(\lm -\rm{\sgn}(\lambda ))}{\lm}+\frac{3}{2\lm} {  c_{F}^2 }\hat m_{F}^{\text{reg}}  -\frac{1}{2 \lm}\frac{(\hat m_{F}^{\text{reg}})^3}{(\lm-\sgn(m_{F}^{\text{reg}}))^2}\nn
& - 3\int_{-\pi}^{\pi} d \alpha \rho(\alpha) \int_{|c_{F}|}^{\infty}dy ~ y
(\ln(1+e^{-y-i\alpha-\nu_F})+\ln(1+e^{-y+i\alpha+\nu_F}))\biggr].
}
\eqref{fefe} agrees precisely with the corresponding result for the 
regular fermion theory (the correct expression is a slight 
generalization of  Eq.(7.20) in \cite{Aharony:2012ns} and Eq.(3.5) in \cite{Jain:2013py}).

Notice that both the leading order finite temperature gap equation and 
the leading order off-shell expression for $v[\rho]$ were independent of 
$g_1, g_2, x_4$ and $x_6$. We conjecture that the same is true for every 
physical quantity in the fermionic scaling limit. For this reason we refer 
to these quantities as spurious or unphysical in the scaling limit under 
consideration.\footnote{In particular we could focus on the special case 
$g_1=0$. In this case $b_4$ is held fixed, rather than being scaled to 
infinity. In other words it is not necessary to scale $b_4$ to infinity 
in order to decouple the bosons.}

\subsection{Bosonic scaling limit}\label{scalarlimitI}

In this subsection we describe the bosonic scaling limit which we 
have already touched upon in the introduction. Our analysis closely 
parallels that of the fermionic scaling limit, so we will keep 
the discussion brief. 

As described in the introduction, the bosonic scaling limit is achieved
by the scaling \eqref{fermionic} with parameters chosen to ensure that 
$c_{B,0}$ is held fixed at ${\tilde c}_{B,0}$ while 
$c_{F,0}$ is taken to infinity. As in the 
previous section, the last condition yields 3 equations that allows us 
to solve for $a_1, a_2$ and $a_3$ and we find% 
\footnote{Note that, scaling limits discussed in this section are not unique. 
For example, one can also obtain critical bosonic theory by scaling  $m_F\rightarrow\infty$ with $b_4 = g_1 m_F^2+g_2 m_F+g_3$ keeping $m_B,x_4,x_6$ fixed.
Following exactly same procedure as described in this section, one can find out $g_1,g_2$ and $g_3$ in terms of parameters which are kept fixed. In this scaling limit as well
 one obtains free energy which is same as that of critical boson.} 
\be\begin{split}\label{bmbs}
&a_1=x_4-\frac{x_4}{(\sgn(m_F)-\lambda )^2},~~a_2=2 m_B^{\text{cri}} 
\lambda  \left(x_4^2 \left(\frac{1}{(\sgn(m_F)-\lambda )^2}-1\right)
+g_1\right),\\
&a_3=\frac{1}{4} (m_B^{\text{cri}})^2 
\left(\lambda ^2 x_4^3 
\left(4-\frac{4}{(\sgn(m_F)-\lambda )^2}\right)-(3 x_6+1)\lm^2+4\right)
+2 m_B^{\text{cri}} g_2 \lambda.
\end{split}\ee
Anticipating that the bosonic scaling limit will lead to the theory of 
a critical boson, we find it convenient to set  
\begin{equation}\label{cr}
{\tilde c}_{B,0}= m_B^{\text{cri}}
\end{equation}
(recall that \eqref{cr} captures the relationship between the pole mass and 
mass parameter for the Lagrangian \eqref{criticalbos}, see \cite{Aharony:2012ns}).

In the limit under consideration ${\cal C}=\frac{|c_F|}{2}$ upto exponential 
corrections. Using this relationship we find that the finite temperature 
gap equation takes the form 
\beal{\label{bselfs}
&c_{F} =\frac{\hat m_{F}-m_{B}^{\text{cri}}x_4\lm}{\sgn(m_F)-\lm}+\frac{b_1}{{\hat m}_{F}}+\frac{b_2}{\hat m_{F}^2}+{\cal O}(\frac{1}{\hat m_{F}})^3\nn
&\cS=\frac{{\hat m}_{B}^{\text{cri}}}{2}+\frac{d_1}{\hat m_{F}}+\frac{d_2}{\hat m_{F}^2}+{\cal O}(\frac{1}{\hat m_{F}})^3\nn
&d_{1} = b_1 \frac{-\sgn(m_F)+\lm}{2x_4\lm},~~d_{2} = b_2 \frac{-\sgn(m_F)+\lm}{2x_4\lm}\nn
&b_1 = -\frac{x_4 (\sgn(m_F)-\lambda ) \left(c_{B}^2-({\hat m}_{B}^{\text{cri}})^2\right)}{2 \lambda  x_4^2 (\lambda -2 \sgn(m_F))-2 g_1 (\lambda -\sgn(m_F))^2}\\
&b_{2} = \frac{x_4 (-\lambda +\sgn(m_F)) (c_{B}^2-({\hat m}_{B}^{\text{cri}})^2) }
{8
   \left(g_1 (\sgn(m_F)-\lambda )^2+\lambda  x_4^2 (2 \sgn(m_F)-\lambda )\right)^2}{\rm B}\nn
&{\rm B}=\lambda  {\hat m}_{B}^{\text{cri}} \left(1+3 x_{6} (\lambda -\sgn(\lm))^2
+\lambda  \left(4 x_{4}^3-1\right) (2 \sgn(\lm)-\lambda )\right)-4(-\lambda +\sgn(m_F))^2 g_2.\notag
}
At leading order, the second of \eqref{bselfs} is simply the gap equation for 
the massive critical boson theory (see Eq.(7.22) of \cite{Aharony:2012ns} and Eq.(3.11) of \cite{Jain:2013py}) provided ${\rm sgn}(m_F)={\rm sgn}(\lambda)$. 
It is easily seen that the inequality \eqref{cb} is obeyed if and only if 
\begin{equation}\label{scaineq}
\sgn(\lm m_F) \geq 0.
\end{equation}
In other words the finite temperature gap equation in the bosonic scaling 
limit agrees with the gap equation of the critical boson theory whenever 
our calculations are reliable.

Substituting into \eqref{free2} we find that the leading order 
off-shell expression for $v[\rho]$ is finite, 
independent of $g_1, g_2, x_4, x_6$, and is given by% 
\footnote{As in the 
previous subsection, it is important to retain subleading corrections in the 
first and second equation in \eqref{bselfs} in order to get this expression 
for $v[\rho]$.} 
\beal{
&v[\rho]\nn
&= {N  \over 6 \pi}  \biggl[ - |c_{B}| ^{3} +\frac{3}{2}\hat m_B^{\text{cri}} c_{B}^2-\frac{1}{2} (\hat m_B^{\text{cri}})^3 \nn
 &+ 3\int_{-\pi}^{\pi}\rho(\alpha)d\alpha\int_{|c_{B}|}^\infty  dy ~ y (\ln \left ( 1- e^{-y-i \alpha-\nu_B}\right) +\ln \left ( 1- e^{-y+i \alpha +\nu_B})\right)\biggr].
}
This is the correct expression for $v[\rho]$ in the critical boson theory 
\eqref{crit}.

\subsection{ The critical scaling limit}

We now turn to a study of the critical scaling limit described in the 
introduction. In this limit we scale $b_4$ to infinity keeping both 
$c_{F,0}$ and $c_{B, 0}$ fixed at the values 
\begin{equation}\begin{split}
c_{F,0}&={\tilde c}_{F,0}={ m_F^{\text{reg}} \over \sgn(\lambda) -\lambda}\\
c_{B,0}&={\tilde c}_{B,0}=m_B^{\text{cri}}.\\
\end{split}
\end{equation}
As described in the introduction, this is achieved by keeping $m_F$ fixed 
but scaling $m_B$ and $b_4$ to infinity according to 
\begin{equation}\label{css} \begin{split}
m_B & \to \infty\\
m_F&=m_{F}^{\text{reg}} +x_4 \lambda m_B^{\text{cri}}\\
b_4&=\frac{1}{2 m_B^{\text{cri}} \lambda} m_B^2 + g_2\\
g_2&=\frac{(m_{F}^{\text{reg}})^2 x_4 \left(\frac{1}{(\sgn(\lambda)-\lambda )^2}-1\right)+\frac{1}{4}  (m_B^{\text{cri}})^2 \left(\lambda ^2 (3 x_6+1)-4\right)}{2 m_B^{\text{cri}}
   \lambda }.\\
\end{split}
\end{equation}

The finite temperature gap equation reduces, in this limit to 
\be\begin{split} \label{ge}
&\cS=\frac{\hat m_B^{\text{cri}}}{2}+\frac{b_1}{\hat m_{B}}+\frac{b_2}{\hat m_{B}^2}+{\cal O}(\frac{1}{\hat m_{B}})^3\\
&b_1 = 0,~~b_2= \frac{1}{2} \hat m_B^{\text{cri}} 
\left( c_{F}^2 x_4-\frac{(\hat m_{F}^{\text{reg}})^2 x_4}{(\sgn(\lambda)-\lambda )^2}-c_{B}^2+ (\hat m_B^{\text{cri}})^2\right).
\end{split}
\ee
Substituting these relations into \eqref{free2} yields a finite 
off-shell expression 
for $v[\rho]$ given by 
\beal{ \label{gege}
&v[\rho]\nn
&= {N  \over 6 \pi}  \biggl[{ | c_F|^3 } \frac{(\lm -\sgn(\lambda) )}{\lm}+\frac{3}{2\lm} {  c_F^2 } \hat m_F^{\text{reg}} 
 -\frac{1}{2 \lm}\frac{(\hat m_F^{\text{reg}})^3}{(\lm-\rm{sgn(m_F^{\text{reg}})})^2} - |c_{B}| ^{3} +\frac{3}{2}\hat m_B^{\text{cri}} c_B^2-\frac{1}{2} (\hat m_B^{\text{cri}})^3\nn
& - 3\int_{-\pi}^{\pi} d \alpha \rho(\alpha) \int_{| c_F|}^{\infty}dy ~ y
(\ln(1+e^{-y-i\alpha-\nu_F})+\ln(1+e^{-y+i\alpha+\nu_F}))   \nn
 &+ 3\int_{-\pi}^{\pi}\rho(\alpha)d\alpha\int_{|c_{B}|}^\infty  dy ~ 
y \left(\ln \left ( 1- e^{-y+i \alpha+\nu_B}\right)+\ln \left ( 1- e^{-y+i \alpha+\nu_B}\right)\right)\biggr].
}
We have independently computed the gap equation and the expression for 
$v[\rho]$ for the system \eqref{crit}; our results agree exactly with those 
presented above at leading order in $\frac{1}{m_B}$
provided that 
\begin{equation}\label{criineq}
\sgn (\lambda) \sgn \left ( 2\lambda\cC+m_F^{\text{reg}} \right )\geq 0.
\end{equation}
It is easily verified that \eqref{criineq} follows as a consequence of 
\eqref{cb}. We conjecture that 
$x_4$ and $x_6$ (which drop out of both the gap equation and the expression for
$v[\rho]$ at leading order) are spurious or unphysical in the 
critical scaling limit.

The system \eqref{crit} itself admits two natural scaling limits. If we take 
the fermionic pole mass of the system to infinity holding the bosonic pole 
mass fixed we are left with a theory of a massive critical boson 
\eqref{criticalbos}. On the 
other hand if we take the bosonic pole mass of the system to infinity holding 
the fermionic pole mass fixed, we are left with the theory of regular 
fermion \eqref{regular}. It is easy to verify that the gap equations 
\eqref{ge} and the expression for $v[\rho]$ \eqref{gege} reduce to 
the correct expressions for the corresponding quantities in the appropriate
limits. 

\subsection{Action of duality on scaling limit}

\subsubsection{Critical boson vs regular fermion}
  
In subsection \ref{fermionlimitI} we described the so called `fermionic' 
scaling limit of the action \eqref{generalaction}. In this limit $x_4$ and 
$x_6$ were held fixed while $m_F$ is scaled to infinity and $b_4$ and 
$m_B^2$ are also scaled to infinity, in a manner determined by the two 
parameters $g_1$ and $g_2$, as described in \eqref{fmbss}.

Applying the duality map \eqref{dmap} to this scaling limit yields a scaling 
expression for the dual quantities $m_B'$, $m_F'$, $b_4'$, $x_4'$ and 
$x_6'$. We have verified that the resultant expressions for the dual quantities 
agree precisely with the bosonic scaling limit of subsection \ref{scalarlimitI}
provided we make the obvious identifications (from \eqref{dmap}) 
\begin{equation}\begin{split}\label{sppar}
x_4'=\frac{1}{x_4}, ~~~x_6'=1+\frac{1-x_6}{x_4^3}, ~~~m_F'=\frac{-m_F}{x_4}
\end{split}
\end{equation}
together with the new identifications 
\be \label{gtrans}
{g'_1} = \frac{1}{x_4} g_1+\frac{3}{4}\frac{1-x_6}{x_4^2},~~~g'_2=-\frac{1}{x_4^2}g_2. 
\ee
($g_1'$ and $g_2'$ are the parameters $g_1$ and $g_2$ that enter the 
bosonic scaling limit \eqref{bmbs}).% 
\footnote{In practice we obtained the transformation relations \eqref{gtrans}
by demanding that the expression for $b_4'$ match the bosonic scaling limit, 
and then verified that the expression for $(m_B')^2$ also matches the bosonic
scaling limit with the same values of $g_1'$ and $g_2'$.

Using this duality map we have also checked that 
\be\begin{split}
c'_{B} &=c_{F}\\
\lm' {\cal C}'&=-\sgn(\lm)\frac{|c_{B}|}{2} +\lm {\cal S}\\
\lm' {\cal S}'&=-\sgn(\lm)\frac{|c_{F}|}{2} +\lm {\cal C}
\end{split}\ee
including all subleading corrections upto terms of order $\frac{1}{m_F^2}$.
}
In other words duality interchanges the fermionic and bosonic scaling limits, 
as claimed in the introduction

\subsubsection{Self duality of the critical scaling limit} 

As in the previous subsection, it is not difficult to examine the action 
of the duality map \eqref{dmap} on the critical scaling limit. We find 
that the dual expressions for all parameters once again fall into the 
critical scaling limit with $x_4$ and $x_6$ transformed according to 
\eqref{dmap} and 
\be
c'_{F,0} = c_{B,0},~~~c'_{B,0} = c_{F,0}.
\ee
This in particular implies that the Lagrangian \eqref{crit} enjoys 
a self duality with the dual interchange of parameters
\begin{equation}\label{critdualt-1} \begin{split}
{m'}_{B}^{\text{cri}} &= -\frac{1}{\lambda -\rm{sgn}(\lm)} m_{F}^{\text{reg}}\\
{m'}_{F}^{\text{reg}}&=-\lambda {m}_{B}^{\text{cri}}\\
\end{split}
\end{equation}
as claimed in the introduction. 

\section{Discussion}

In this paper we have presented exact expressions for the pole masses 
and thermal free energy of the class of quantum field theories 
\eqref{generalaction} parameterized by 7 continuous and two discrete 
parameters, in the large $N$ limit.  Our results allowed us to conjecture 
that the class of theories \eqref{generalaction} enjoys invariance under 
a strong weak  coupling duality transformation under which the 
parameters of the action transform according to \eqref{dmap}. We have also 
identified several interesting scaling limits of the action 
\eqref{generalaction} and demonstrated that duality interchanges these limits. 

The duality \eqref{dmap} proposed in this paper reduces to the supersymmetric
Giveon Kutasov duality in one limit and to the `bosonization' duality between 
Chern Simons coupled regular fermions and critical bosons in another limit. 
In other words the nonsupersymmetric bosonization duality may be regarded 
as a deformation of the supersymmetric Giveon Kutasov duality% 
\footnote{ Note that, for instance, the inequality \eqref{cb} is obeyed, in the 
Fermionic scaling limit, provided that the \eqref{fermineq} is obeyed. 
We see no obstruction to deforming the ${\cal N}=2$ superconformal theory 
to the fermionic scaling limit s.t. the inequality \eqref{cb} is obeyed 
along the entire path. We thank O. Aharony for discussions on this issue. }
As the 
supersymmetric duality is believed to be true at finite $N$, the 
analysis presented in our paper suggests that the nonsupersymmetric 
bosonization duality is also valid at finite $N$.\footnote{
We have not succeeded in reproducing 
the conjectured duality between critical fermions and regular bosons as 
a limit of the duality \eqref{dmap}; it would be interesting to see if this 
is possible.}

The general field theory \eqref{generalaction} is always renormalizable (and 
so well defined) in in the strict large $N$ limit studied in this paper. 
As we have emphasized in the introduction, however, it could well be that 
only a subclass of these theories is renormalizable (so well defined)
 at finite $N$. As we have explained in the introduction, one way of 
analyzing this question would be to compute the scaling dimensions of the 
operators multiplying $x_4$, $x_6$, $y_4'$ and $y_4''$, about the 
${\cal N}=2$ fixed point, at first subleading order in the $\frac{1}{N}$ 
expansion. It would be very interesting (and may be possible) to carry 
out the relevant computations.

As we have explained in the text and in Appendix \ref{eea}, 
the gap equations and thermal free energies presented in this paper 
enjoy invariance under duality only provided certain inequalities are obeyed. 
We have interpreted this fact to suggest that the bosonic field in 
the problem condenses when these inequalities are not obeyed 
(see closely related remarks in \cite{Aharony:2012ns}). If this 
conjecture is correct  it would be 
very interesting to honestly compute the free energy of our system in the 
condensed phase, and to verify duality of the free energy in the full 
range of parameter space.\footnote{As we have emphasized in Appendix \ref{bosecondensation} (see 
also \cite{Aharony:2012ns}) a very similar issue arises even in the simplest example 
of this bosonization duality.} This issue is very interesting 
from the physical viewpoint: Bose condensation is very 
bosonic behaviour, so it is particularly interesting to see how it 
interplays with bose-fermi duality.

It should be straightforward (if potentially messy) to repeat the analysis 
of this paper to the theory with two fundamental scalars and two fundamental
fermions. The class of theories with this field content includes 
${\cal N}=2$ theories with one fundamental and one antifundamental chiral 
multiplet and an  ${\cal N}=3$ theory. The conjectured Giveon Kutasov duality
for these theories has a new element; the dual theory includes a singlet 
`Lagrange multiplier' type chiral field. It would be interesting to 
investigate how this new field impacts the general
nonsupersymmetric.

It would 
be interesting to identify the Vasiliev duals of the general class of 
theories \eqref{generalaction}. As all these theories may be obtained 
from the ${\cal N}=2$ theory via relevant and marginal deformations by 
local single and multitrace operators, it follows that it should be 
possible to obtain the Vasiliev duals of these theories from that of the 
${\cal N}=2$ theory described in \cite{Chang:2012kt} by a suitable deformation of 
boundary conditions. This exercise is of interest, as we expect the 
nontrivial strong weak coupling duality \eqref{dmap} to be manifest 
from the Vasiliev point of view.

It would be very interesting to study the exact S 
matrix for scattering of (say) four partons (four scalars, four fermions, or 
two scalars with two fermions) in the large $N$ limit. The graphs that 
determine this quantity are summed by the solution to an integral equation, 
whose form was determined in \cite{Giombi:2011kc} in the special case of the 
fermionic theory. It is possible that a careful study of the generalization 
of these integral equations can be used to prove the invariance of this 
S matrix under duality. We hope to return 
to this question in the future. 

\acknowledgments
We would like to thank S. Bhattacharyya,  T. Sharma and T. Takimi for 
initial collaboration on this project. We would also like to thank 
O. Aharony, S.Banerjee, S. Trivedi and S. Wadia  for useful discussions, and 
C. M. Chang, S. Giombi, S. Wadia,  X. Yin and especially 
O. Aharony for useful comments on our manuscript.   The work of SM is 
supported by a 
Swarnajayanti Fellowship. We would all also like to  acknowledge our debt 
to the people of India for their generous and steady support to research 
in the basic sciences.

%%%%%%%%%%%%%%%%%%%%%%%%%%%%%%%%%%%%%%%%%%%%%%%
\appendix
\section{Details of the computation}

\subsection{Exact effective action} 
\label{eea}

In this appendix, we illustrate how to derive the free energy \eqref{free1} from the general 
action \eqref{generalaction}. The method is the same as that used in \cite{Giombi:2011kc,Jain:2012qi}.

The first thing to do is to integrate out gauge field from the original action \eqref{generalaction} by taking the light-cone gauge \cite{Giombi:2011kc}. 
After integrating out gauge field the action becomes
\beal{
&S =  \int \frac{d^3q}{(2 \pi)^3}  \left( \bar\phi(-q) (\wt q^2 +m_B^2) \phi(q)  +\bar{\psi}(-q) (i\gamma^\mu \wt q_\mu +m_F) \psi(q)    \right) \nonumber \\
&+N \int  \frac{d^3P}{(2 \pi)^3} \frac{d^3q_1}{(2 \pi)^3} \frac{d^3q_2}{(2 \pi)^3} 
 C_1(P,q_1,q_2)  \chi(P,q_1) \chi(-P,q_2) \nn
&+ N \int \frac{d^3P_1}{(2 \pi)^3} \frac{d^3P_2}{(2 \pi)^3}  \frac{d^3q_1}{(2 \pi)^3} \frac{d^3q_2}{(2 \pi)^3}\frac{d^3q_3}{(2 \pi)^3} C_2(P_1,P_2,q_1,q_2,q_3)
\chi(P_1,q_1)\chi(P_2,q_2)\chi(-P_1-P_2,q_3)\nn
&+N \int \frac{d^3P}{(2 \pi)^3} \frac{d^3q_1}{(2 \pi)^3} \frac{d^3q_2}{(2 \pi)^3}~
\frac{8 \pi iN}{ k (q_1-q_2)_{-}} \xi_-(P,q_1) \xi_I(-P,q_2)\nn
&+ N\int \frac{d^3P}{(2 \pi)^3} \frac{d^3q_1}{(2 \pi)^3} \frac{d^3q_2}{(2 \pi)^3}
{ 8\pi N x_4\over k} \chi(P,q_1) \xi_I(-P,q_2) + \cdots,
}
where   
\beal{
&\chi(P,q)=\frac{1}{N}\bar{\phi}(\frac{P}{2}-q) \phi(\frac{P}{2}+q), \quad \\
&\xi_I(P,q) = \frac{1}{2N}\bar{\psi}(\frac{P}{2}-q) \psi(\frac{P}{2}+q), \quad
\xi_-(P,q) = \frac{1}{2N}\bar{\psi}(\frac{P}{2}-q) \gamma_-\psi(\frac{P}{2}+q).
\label{xi}\\
&C_1(P,q_1,q_2)={2 \pi i N \over k} \frac{((-P+q_1+ q_2)_{3}+\mu_B) (P+q_1+q_2)_{-}}{(q_1-q_2)_{-}}+ {4\pi N b_4\over k}, \\
&C_2(P_1,P_2,q_1,q_2,q_3)=
{4 \pi^2 N^2 \over k^2} \frac{( P_1 - P_2+2q_1+2q_2)_{-} (P_1+2P_2+ 2q_2 + 2q_3)_{-}}{(P_1+P_2+ 2q_1-2q_2)_{-} (P_1-2q_2+2q_3)_{-}} \nn 
&\qquad\qquad\qquad\qquad~~~~  + {4 \pi^2 N^2 x_6 \over k^2}.
}
and the ellipsis contains $(\bar\psi \phi)$ and $(\bar\phi\psi)$ terms containing $y_4', y_4''$ couplings as well as the $1/N$ correction terms.

The next thing is to assume the translational invariance for the fields to study vacuum structure of this system. 
\be
\chi (P, q ) = (2\pi)^3 \delta^3(P) \chi(q) \quad 
\xi_I(P,q) = (2\pi)^3 \delta^3(P) \xi_I(q) , \quad
\xi_-(P,q) = (2\pi)^3 \delta^3(P) \xi_-(q) .
\ee
Then the action reduces to 
\beal{ S =&  \int \frac{d^3q}{(2 \pi)^3}  \left( \bar\phi(-q) (\wt q^2 +m_B^2) \phi(q)  +\bar{\psi}(-q) (i\gamma^\mu \wt q_\mu +m_F) \psi(q)   \right) \nonumber \\
&+N V \int  \frac{d^3P}{(2 \pi)^3} \frac{d^3q_1}{(2 \pi)^3} \frac{d^3q_2}{(2 \pi)^3} { 4\pi N b_4\over k}  \chi(q_1) \chi(q_2) \nn
&+NV \int \frac{d^3q_1}{(2 \pi)^3}\frac{d^3q_2}{(2 \pi)^3}\frac{d^3q_3}{(2 \pi)^3}
 C_2(0,0,q_1,q_2,q_3) \chi(q_1) \chi(q_2)\chi(q_3) \nn
&+N V \int \frac{d^3q_1}{(2 \pi)^3} \frac{d^3q_2}{(2 \pi)^3}~
\frac{8 \pi iN}{ k (q_1-q_2)_{-}} \xi_-(q_1) \xi_I(q_2)\nn
&+ NV \int \frac{d^3q_1}{(2 \pi)^3} \frac{d^3q_2}{(2 \pi)^3}
{ 8\pi N x_4\over k} \chi(q_1) \xi_I(q_2),
\label{S1}
}
in the leading of large $N$, where $V = (2\pi)^3 \delta(0)$.  
Note that the couplings $y'_4, y_4''$ have gone away from the action at this stage. 

The third thing is to introduce auxiliary fields and rewrite the interaction terms in terms of them.
We add the following terms into the action \eqref{S1}. 
\beal{
& - NV\int \frac{d^3 q}{(2 \pi)^3}\( \Sigma_B( q) (\alpha_B(q) - \chi(q)) +  2\Sigma_F^I( q) (\alpha_F{}_I(q) - \xi_I(q)) + 2\Sigma_F^-( q) (\alpha_F{}_-(q) - \xi_-(q)) \) \nn
&-S_{int}(\chi, \xi_I, \xi_-) + S_{int}(\alpha_B, \alpha_{FI}, \alpha_{F-})
}
where $S_{int}(\chi, \xi_I, \xi_-)$ represents the interaction terms. 
Then the action including these terms denoted by $\tilde S$ becomes 
\beal{
\tilde S 
=&   \int \frac{d^3q}{(2 \pi)^3}  \bar\phi(-q) \left( \wt  q^2 +m_B^2 + \Sigma_B(q)  \right)  \phi(q)  
+\bar{\psi}(-q) (i\gamma^\mu \wt q_\mu +m_F + \Sigma_F(q)) \psi(q) \nn
&+N V \int  \frac{d^3P}{(2 \pi)^3} \frac{d^3q_1}{(2 \pi)^3} \frac{d^3q_2}{(2 \pi)^3}  { 4\pi N b_4\over k}  \alpha_B(q_1) \alpha_B(q_2) \nn
&+NV \int \frac{d^3q_1}{(2 \pi)^3}\frac{d^3q_2}{(2 \pi)^3}\frac{d^3q_3}{(2 \pi)^3}
C_2(0,0,q_1,q_2,q_3) \alpha_B(q_1) \alpha_B(q_2)\alpha_B(q_3) \nn 
&+N V \int \frac{d^3q_1}{(2 \pi)^3} \frac{d^3q_2}{(2 \pi)^3}~
\frac{8 \pi iN}{ k (q_1-q_2)_{-}} \alpha_F{}_-(q_1) \alpha_F{}_I(q_2)\nn
&+ NV \int \frac{d^3q_1}{(2 \pi)^3} \frac{d^3q_2}{(2 \pi)^3}
{ 8\pi N x_4\over k} \alpha_B(q_1) \alpha_F{}_I(q_2) \nn
& - NV\int \frac{d^3 q}{(2 \pi)^3} \Sigma_B( q) \alpha_B(q) +  2\Sigma_F^I( q) \alpha_F{}_I(q) 
+ 2\Sigma_F^-( q) \alpha_F{}_-(q) 
}
where  $\Sigma_F = \Sigma^+_{F} \gamma_+  +\Sigma^-_{F}\gamma_{-}+\Sigma^3_{F}\gamma_{3}+\Sigma^I_{F} I$
with $\Sigma^+_{F} = \Sigma^3_{F}=0$.
Since this is quadratic in terms of the bosonic and fermionic fields, 
one can integrating them out by Gaussian integration, which gives 
\beal{
\tilde S &= NV \biggl\{ \int \frac{d^3q}{(2 \pi)^3} \left( \log (\wt q^2 +m_B^2+\Sigma_B(q) )
-  \log ( i \gamma^\mu \wt q_\mu +m_F+ \Sigma_{F}(q)) \right) \nn
&+ \int  \frac{d^3q_1}{(2 \pi)^3} \frac{d^3q_2}{(2 \pi)^3} { 4\pi N b_4\over k}  \alpha_B(q_1) \alpha_B(q_2) \nn
&+\int \frac{d^3q_1}{(2 \pi)^3} \frac{d^3q_2}{(2 \pi)^3}\frac{d^3q_3}{(2 \pi)^3} C_2(0,0,q_1,q_2,q_3)
\alpha_B(q_1)\alpha_B(q_2)\alpha_B(q_3) \nn
&+\int  \frac{d^3q_1}{(2 \pi)^3} \frac{d^3q_2}{(2 \pi)^3}~
\frac{8 \pi i \lambda}{  (q_1-q_2)_{-}} \alpha_{F,-}(q_1) \alpha_{F,I}(q_2) + \int \frac{d^3q_1}{(2 \pi)^3} \frac{d^3q_2}{(2 \pi)^3}
8\pi \lambda x_4  \alpha_B(q_1) \alpha_{F,I}(q_2) \nn
&+ \int  \frac{d^3q}{(2 \pi)^3} 
\biggl(-\Sigma_B (q) \alpha_B(q) -2 \Sigma_{F,+}(q) \alpha_{F,-}(q)-2 \Sigma_{F,I}(q) \alpha_{F,I}(q)
\biggr) \biggr\}.
}

Equations of motion for $\Sigma_B, \Sigma_F$ are 
\beal{
\label{alphab}
\alpha_B(q)=& {1\over \wt q^2 +m_B^2+ \Sigma_B(q)}   \\
\alpha_F(q)=& -  {1\over  i\gamma^\mu \wt q_\mu +m_F+\Sigma_F(q) }  
\label{alphaf}
}
where $\alpha_F = \alpha_{F,+} \gamma^+ +\alpha_{F,-}\gamma^{-}+\alpha_{F,3}\gamma^{3}+\alpha_{F,I}I$  with $\alpha_{F,+} = \alpha_{F,3}=0$.

Equations of motion for $\alpha_B, \alpha_F$ are 
\beal{ 
{\Sigma}_{F,+}(p)=&{4 \pi i \lambda}\int \frac{d^3q}{(2\pi)^3}\frac{1}{(p-q)_{-} } \alpha_{F,I}(q),
\label{sigmaf+}\\
{\Sigma}_{F,I}(p)=&-{4\pi i \lambda}\int \frac{d^3q}{(2\pi)^3} \frac{1}{(p-q)_{-}} \alpha_{F,-}(q) 
+4\pi \lambda x_{4} \int \frac{d^{3}q}{(2\pi)^3} \alpha_B(q),
\label{sigmafI}\\ 
{\Sigma}_{B}(p) =&\int \frac{d^{3}q}{(2\pi)^3}\frac{d^{3}q'}{(2\pi)^3} \Big\lbrack C_2(p,q,q')+C_2(q,p,q')+ C_2(q,q',p)\Big\rbrack \alpha_B(q) \alpha_B(q') \nn
&+8 \pi \lambda x_{4} \int \frac{\cD^3 q}{(2\pi)^3} \alpha_F^I (q) + { 8 \pi N b_4\over k} \int  \frac{d^3q}{(2 \pi)^3}  \alpha_B(q).
\label{sigmasf}
}
By using these equations of motion the effective action can be simplified on shell as follows. 
\beal{
\tilde S &=NV\biggl[ \int \frac{d^3q}{(2 \pi)^3}   \biggl\{ \log (\wt q^2 +\wt\Sigma_B(q) )
-\log ( i \gamma^\mu \wt q_\mu +\wt \Sigma_{F}(q)) \nn
&-\frac{2}{3} \Sigma_B(q)  \( {1\over \wt q^2 + \wt\Sigma_B(q)}  \)
+ \frac{1}{2}\tr\left({\Sigma}_{F}(q)\frac{1}{i \gamma^\mu \wt q_\mu +\wt {\Sigma}_F(q)}\right) 
\biggr\} + {4\pi \lambda b_4 \over 3} \( \int  \frac{d^3q}{(2 \pi)^3}{1\over \wt q^2 + \wt\Sigma_B(q)}\)^2 \nn
&- \frac{2\pi \lambda x_{4}}{3}
\left( \int\frac{d^{3}q}{(2\pi)^3}\frac{1}{\wt q^{2}+\wt{\Sigma}_{B}(q)} \right)
\left( \int\frac{d^{3}p}{(2\pi)^3} \tr\frac{1}{i \gamma^\mu \wt p_\mu + \wt{\Sigma}_F(p) }\right)
\biggr]
}
where we define
\be
\wt \Sigma_B=\Sigma_B+m_B^2, \quad \wt\Sigma_F=\Sigma_F+m_F.
\ee
% \subsection{Zerotemperature}
% \be\begin{split}
% &c_{B,0}=-\frac{4 \lambda  \left(b_4 (\lambda -1)^2-(\lambda -2) \lambda  m_F x_4^2+2(1-\lambda) \sqrt{X1}\right)}
% {-8 \lambda +\lambda ^4 \left(4 x_4^3-3 x_6-1\right)+\lambda ^3 \left(-8 x_4^3+6 x_6+2\right)-3 \lambda ^2 (x_6-1)+4}\nonumber\\
% \end{split}\ee
% where 
% \be\begin{split}
% &X1= \lambda  \left(4 b_4^2 \lambda +(\lambda -2) \left(4 b_4^2 \lambda ^2+m_{f} x_4 \left(\lambda ^2 (-8 b_4 x_4+3 m_{f}
%    x_6+m_{f})-4 m_{f}\right)\right)\right)\nonumber\\
% &+m_{B}^2 \left(\lambda  \left(\lambda  \left((\lambda -2) \lambda  \left(4 x_4^3-1\right)-3
%    (\lambda -1)^2 x_6+3\right)-8\right)+4\right)
% \end{split}\ee
% In the  $b_4\rightarrow\infty$ we get
% \be
% c_{B,0}=\frac{ m_{B}^2}{2 b_4  \lambda }+{\cal O}(\frac{1}{b_4}).
% \ee
% keeping $\frac{m_{B}^2}{b_4},~~\frac{m_{f}^2}{b_4}$ fixed.
This is the exact effective action derived from the general action \eqref{generalaction} 
in the leading of the large $N$ limit.

\subsection{Exact thermal free energy}
\label{freeenergy}

Let us move on to computation at finite temperature taking into account holonomy effect along the 
thermal circle whose circumference is identified with the inverse temperature.

From the dimensional grounds, we set the ansatz for the saddle point equations as follows.  
\be
\wt\Sigma_B = c_B^2 T^2, \quad 
\wt\Sigma_{F,I}(p) = f(\hat p) p_s, \quad 
\wt\Sigma_{F,+}(p) =i g(\hat p) p_+. 
\label{ansatzsigmaf}
\ee
Since the saddle point equations for $\Sigma_F$, \eqref{sigmaf+}, \eqref{sigmafI}, are 
the same as those in \cite{Aharony:2012ns}, 
and the computation is the same as in \cite{Jain:2012qi}, 
we just mention the results here.  
$f(\hat p)$ and $g(\hat p)$ satisfying \eqref{sigmaf+}, \eqref{sigmafI} 
are given by \eqref{sigmafg} and \eqref{girfreesquare}.

On the other hand, a new term containing $b_4$ coupling appears in the saddle equation for $\Sigma_B$, 
\eqref{sigmasf}, so we will demonstrate how to solve it. 
From \eqref{sigmasf}, we can show $\partial_{q_+}\Sigma_B(q)=0$. 
By assuming the rotational invariance for $\Sigma_B(q)$ with respect to 2-plane, 
it becomes constant. $\Sigma_B(q) = \Sigma_B$. 
Therefore we can compute the right-hand side of \eqref{sigmasf} as
\begin{align}
\Sigma_B 
&= \int \frac{\cD^3q_2}{(2 \pi)^3}\frac{\cD^3q_3}{(2 \pi)^3}
( C_2(0,q_2,q_3)+ C_2(q_2,0,q_3)+ C_2(q_3,q_2,0)) \alpha_B(q_2)\alpha_B(q_3)\nn
&+8 \pi \lambda x_{4} \int \frac{\cD^3 q}{(2\pi)^3} \alpha_F^I (q) 
+{ 8 \pi \lambda b_4 } \int  \frac{\cD^3q}{(2 \pi)^3} \alpha_B (q)\nn
&= (2 \pi)^2  \lambda^2 \(1 + 3 x_6 \)  \(\int \frac{\cD^3q}{(2 \pi)^3} \alpha_B(q) \) ^2 
+8 \pi \lambda x_{4} \int \frac{\cD^3 q}{(2\pi)^3} \alpha_F^I (q)
+8 \pi \lambda b_{4}\int  \frac{\cD^3q}{(2 \pi)^3} \alpha_B (q) \notag
\end{align}
where we denoted by $\cD^3 q$ the integration measure in the Fourier space in $\bR^2\times \bS^1$ , 
which is also used in \cite{Aharony:2012ns}. 
Thus we obtain
\begin{align}
% \Sigma_B 
% &= (2 \pi \lambda_{\text{eff}})^2 \(  - { 1 \over 2 \pi \beta} \cS \) ^2 + 2N\lambda^b_4 \(  - { 1 \over 2 \pi \beta} \cS \)+
% 8\pi \lambda x_4 {1 \over \pi \beta^2} \( \half \lambda\cC -x_4\lambda \cS +\half \hat m_F \) \cC  \nn
% \Leftrightarrow \quad 
c_B^2 
&= \lambda^2 \(1 + 3 x_6 \) \cS^2 -4\lambda \hat b_4 \cS + 4 x_4 \(  \lambda^2\cC^2 - 2 x_4 \lambda^2 \cS \cC +\hat m_F  \lambda \cC \) +\hat m_B^2
\label{bosfreeappendix}
\end{align}
where we used \eqref{ansatzsigmaf} and computed the right-hand side in the following manner. 
\beal{
\int \frac{\cD^3q}{(2 \pi)^3} {1\over \wt q^2 + \wt \Sigma_B}  =&  - { 1 \over 2 \pi \beta} \cS, \nn
\int \frac{\cD^3q}{(2 \pi)^3} \alpha_F^I(q) =& {1 \over \pi \beta^2} \( \half  \lambda \cC -x_4 \lambda  \cS  +\half \hat m_F \) \cC.
}

By using these solutions of the saddle point equations, 
the effective action of the system on $\bR^2\times \bS^1$ can be simplified as 
\beal{
 { 6 \pi \tilde S \over NV_2T^2} 
=& - |c_B|^3 + 2 (c_B^2 - \hat m_B^2 )\cS   + 3 \int d\alpha \rho(\alpha) \int_{|c_B|} ^\infty dy y \( \log (1 - e^{-y-i\alpha -\nu_B}) +\log (1 - e^{-y+i\alpha +\nu_B}) \) \nn
& + |c_F|^3 - 2 (c_F^2 - \hat m_F^2) \cC
- 3  \int d\alpha \rho(\alpha) \int_{ |c_F|} ^\infty dy y \( \log (1 + e^{-y-i\alpha -\nu_F }) + \log (1 + e^{-y + i\alpha + \nu_F }) \) \nn
&+ 2 \lambda \hat b_4 \cS^2 + 2 \lambda \hat m_F \cC^2 - 4\lambda x_4\hat m_F\cS\cC .
\label{free0}
}

To determine the free energy (or grand-canonical potential) which respects the duality, 
one has to be careful about the normalization thereof. 
The normalization is such that 
the free energy goes to zero when the temperature does \cite{Aharony:2012ns}. 
As a result we determine the thermal free energy respecting the duality as follows. 
\beal{
\!\!\! { 6 \pi v[\rho] \over N} 
=& -| c_B|^3 + 2 (c_B^2 - \hat m_B^2 )\cS   + 3 \int d\alpha \rho(\alpha) \int_{|c_B|} ^\infty dy y \( \log (1 - e^{-y-i\alpha -\nu_B}) +\log (1 - e^{-y+i\alpha +\nu_B}) \) \nn
& + |c_F|^3 - 2 (c_F^2 - \hat m_F^2) \cC
- 3  \int d\alpha \rho(\alpha) \int_{ |c_F|} ^\infty dy y \( \log (1 + e^{-y-i\alpha -\nu_F }) + \log (1 + e^{-y + i\alpha + \nu_F }) \) \nn
&+ 2 \lambda \hat b_4 \cS^2 + 2 \lambda \hat m_F \cC^2 - 4\lambda x_4\hat m_F\cS\cC \nn
&+|\hat c_{B,0}|^3+|\hat c_{B,0}| \hat m_{B}^2
-|\hat c_{F,0}|^3 - |\hat c_{F,0}| \hat  m_{F}^2-\frac{\lambda }{2} \hat b_4\hat c_{B,0}^2 
-\frac{\lambda}{2}\hat m_F \hat c_{F,0}^2   + x_4 \lambda \hat m_{F}  |\hat c_{B,0}| |\hat  c_{F,0}|
}
where $c_{F,0}, c_{B,0}$ are the pole mass of scalar and fermion fields in the zero temperature 
determined by \eqref{ztm}, \eqref{fpm}.
% \beal{
% c_{F,0}^2 =& (\lambda(c_{B,0}-x_4 c_{F,0})+m_F)^2 
% \label{cf0} \\
% c_{B,0}^2 =& \lambda^2 (1+3x_6) {c_{B,0}^2 \over 4} -2\lambda  b_4 {c_{B,0} }  +  x_4 { \lambda (-\lambda + 2s) \over  (\lambda -s )^2} ( - m_F + x_4 \lambda c_{B,0} )^2 + m_B^2. 
% \label{cb0}
% }
\subsection{Explicit expression for off-shell thermal free energy}
\label{off-shellfreeenergy}

As explained in the main text, in order to obtain the off-shell free energy 
we use the gap equations to solve for $\cS, \cC$ as functions of $c_B$ and 
$c_F$ and plug the resultant expressions into \eqref{free2}. 
Using \eqref{girfreesquare} we obtain
\begin{equation}\label{cCf}
 \cC=\frac{\sgn(\text{X})  |c_F| - \hat m_F}{2 \lambda }+ \cS x_4,
\end{equation}
where $\text{X} = 2\lm (\cC-x_4 \cS)+\hat m_F.$
Substituting above in \eqref{bosfree} and solving for $\cS$ gives (note that, here we have chosen particular branch which reduces that of 
regular bosonic theory when we set $x_4,b_4,m_B$ to zero.) 
\begin{equation}\label{cSb}
\cS=\frac{\frac{1}{2} \sqrt{\left(4 \hat b_4 \lambda - 4 \lambda  \hat m_F
   x_4^2\right)^2-4 \lambda ^2 \left(4 x_4^3-3 x_6-1\right)
   \left(x_4 \left(\hat m_F^2-
   c_F^2\right)+c_B^2-\hat m_B^2\right)}-2 \hat b_4 \lambda +2 \lambda \hat m_F x_4^2}{ \lambda^2 \left(4 x_4^3 -3 x_6 - 1 \right)}
\end{equation}
Plugging $\cC,~\cS$ using \eqref{cSb},\eqref{cCf} in \eqref{free1} we obtain
\be\label{off-shellfreeenergy1}
\begin{split}
&-\frac{6\pi v[\rho]}{N}=|c_B|^3-|c_F|^3+\frac{2 \left(c_B^2-\hat m_{B}^2\right) \left(2 b_4 \lambda -2 \lambda  \hat m_F x_4^2-\frac{Z}{2}\right)}{\lambda ^2 \left(4 x_4^3-3
   x_6-1\right)}-\frac{2 \hat b_4 \left(2 \hat b_4 \lambda -2 \lambda  \hat m_F x_4^2-\frac{Z}{2}\right)^2}{\lambda ^3 \left(-4
   x_4^3+3 x_6+1\right)^2}\\
&+\frac{\left(c_F^2-\hat m_F^2\right) \left(\lambda  (\sgn(\text{X}) |c_F|-\hat m_F)+\frac{2 x_4 \left(-2 \hat b_4 \lambda +2 \lambda  \hat m_F
   x_4^2+\frac{Z}{2}\right)}{4 x_4^3-3 x_6-1}\right)}{\lambda ^2}\\
&-\frac{\hat m_F \left(\lambda  (\sgn(\text{X}) |c_F|-\hat m_F)+\frac{2
   x_4 \left(-2 \hat b_4 \lambda +2 \lambda  \hat m_F x_4^2+\frac{Z}{2}\right)}{4 x_4^3-3 x_6-1}\right)^2}{2 \lambda ^3}\\
&+\frac{2 \hat m_F x_4 \left(-2 \hat b_4 \lambda +2 \lambda  \hat m_F x_4^2+\frac{Z}{2}\right) \left(\sgn(\text{X}) |c_F| \lambda (-1+4 x_4^3-3 x_6) 
   -4 \hat b_4 \lambda  x_4+\lambda  \hat m_F+3 \lambda  \hat m_F x_6+Z
   x_4\right)}{\lambda ^3 \left(-4 x_4^3+3 x_6+1\right)^2}\\
&- 3 \int d\alpha \rho(\alpha) \int_{ |c_B|} ^\infty dy y \left( \log (1 - e^{-y-i\alpha -\nu_B}) +\log (1 - e^{-y+i\alpha +\nu_B}) \right) \\
& + 3  \int d\alpha \rho(\alpha) \int_{ |c_F|} ^\infty dy y \left( \log (1 + e^{-y-i\alpha -\nu_F }) + \log (1 + e^{-y + i\alpha + \nu_F }) \right) \\
&-\left(|\hat c_{B,0}|^3-|\hat c_{F,0}|^3-\frac{1}{2} \hat b_4 \hat c_{B,0}^2 \lambda ^2-\frac{1}{2}\hat c_{F,0}^2 \lambda ^2
  \hat m_{F}+|\hat c_{B,0}| \lambda 
   \hat m_{B}^2-|\hat c_{F,0}| \lambda  \hat m_{F}^2 +|\hat c_{B,0}|
   |\hat c_{F,0}| \lambda ^2 \hat m_{F} x_4\right)
\end{split}
\ee
where 
\be
Z=\sqrt{\left(4 \hat b_4 \lambda -4 \lambda  \hat m_F x_4^2\right)^2-4 \lambda ^2 \left(4 x_4^3-3 x_6-1\right)
   \left(c_B^2+x_4 \left(\hat m_F^2-c_F^2\right)-\hat m_B^2\right)}.
\ee

One can check that differentiating free energy with $c_F$ and $c_B$ and using \eqref{cSb} and \eqref{cCf} we get
\be
\partial_{c_B}F=0,~~\partial_{c_F}F=0.
\ee
It is easy to check with this form of off-shell free energy that, under the duality transformations \eqref{dmap}, \eqref{off-shellfreeenergy1} is invariant provided $\sgn(\text{X})=\sgn(\lm)$.
\section{Bose Condensation?}\label{bosecondensation}

In this section we study the regular massive fermion coupled to a Chern 
Simons gauge field \eqref{regular}. We present a detailed analysis
of the following question: in what region of parameter space of this 
theory does its gap equation reproduce the naive, uncondensed gap 
equation of its putative dual, the critical boson theory \eqref{criticalbos} ?

In the complex lightcone gauge 
employed in this paper the fermion propagator is taken to be same as in \cite{Giombi:2011kc}, see Eq.(2.2) and Eq.(2.12).
% \begin{equation}\label{propagator}
% \frac{1}{i \gamma^\mu p_\mu +\Sigma(p)}
% \end{equation}
% where 
% $$\Sigma(p)= f p_s + i g p_+$$
The solution to the gap equation is\footnote{We have used the convention
$$\alpha \in (-\pi, \pi]$$ 
and the logarithm in all formulae above and subsequently is taken to have a 
branch cut along the negative real axis.}
 
\begin{equation}\label{selfferc}\begin{split}
 f(\hat p_s)&=\frac{1}{\hat p_s}\hat m_{F}^{\text{reg}}+\frac{\lambda}{\hat p_s}\int_{-\pi}^{\pi} d\alpha~\rho(\alpha)\left(\ln 2\cosh(\frac{\sqrt{\hat p_s^2+c_{F}^2}+i \alpha +\nu_F}{2})+\ln 2\cosh(\frac{\sqrt{\hat p_s^2+c_{F}^2}-i \alpha -\nu_F}{2})\right),\\
g(\hat p_s)&=\frac{c_{F}^2}{\hat p_s^2}-f(\hat p_s)^2,\\
c_{F}^2&= 
\left( \hat m_{F}^{\text{reg}} + \lambda \int_{-\pi}^{\pi} d\alpha~\rho(\alpha)\left(\ln 2\cosh(\frac{c_{F}+i \alpha + \nu_F}{2})+\ln 2\cosh(\frac{c_{F}-i \alpha -\nu_F }{2})\right) 
\right)^2.
\end{split}
\end{equation}

Note that the propagator  is independent
of the sign of ${c_{F}}$ (it turns out that the same is true for the free 
energy). In other words the apparent ambiguity of 
sign in the third equation of \eqref{selfferc} is a fake. We can choose 
either branch for the solution of this equation. We choose to work with
 the branch 
\be \label{gap}
c_{F}= \hat m_{F}^{\text{reg}} + \lambda \int_{-\pi}^{\pi} d\alpha~\rho(\alpha)\left(\ln 2\cosh(\frac{c_{F}+\nu_F+i \alpha}{2})
+\ln 2\cosh(\frac{c_{F}-\nu_F-i \alpha}{2})\right).
\ee

\subsection{Space of solutions of \eqref{selfferc}}
 
Let us define 
\be \label{cdef}
{\cal C}(c_F)= \frac{1}{2} \int_{-\pi}^{\pi} d\alpha~\rho(\alpha)
\left(\ln 2\cosh(\frac{ c_F+i \alpha}{2})+\ln 2\cosh(
\frac{ c_F-i \alpha}{2})\right).
\ee
The  last of \eqref{selfferc} may be rewritten as 
$${c_{F}}^2=\left( \hat m_{F}^{\text{reg}} + \lambda {\cal C}({c_{F}}+ \nu_F , \lambda) 
+  \lambda {\cal C}({c_{F}}- \nu_F , \lambda) \right)^2.$$
We assume throughout that $\rho(\alpha)$ is an even function of $\alpha$.
With this assumption it is easily verified that  
$\lambda {\cal C}(c_{F})$ is an even 
function of $x$ but an odd function of $\lambda$. The function 
${\cal C}$ is always real and positive and asymptotes to $\frac{ |c_{F}|}{2}$ 
at large values of $c_{F}$. As ${\cal C}(c_{F})$ is an even function of 
$c_{F}$ its  derivative w.r.t $c_{F}$ vanishes at $c_{F}=0$. 

Using these properties of the function ${\cal C}(c_{F})$ and drawing a 
few graphs, it is easy to convince oneself that \eqref{gap} has a unique solution
for every value of $\hat m_{F}^{\text{reg}}$, $\nu_F$, and $\lambda \in (-1, 1]$. 
The solution for ${c_{F}}$ has the same sign as $\hat m_{F}^{\text{reg}} +\lm {\cal C}(\nu_F)$. 
The solution to the equation \eqref{gap} varies continuously as a function 
$\hat m_{F}^{\text{reg}}$ and $\nu_F$. It is an analytic function of these variables 
provided that ${\cal C}(c_F)$ is an analytic function of $c_F$. 
${\cal C}(c_F)$ is in fact an analytic function provided $\rho(\alpha)$ vanishes
in an interval around $\alpha=\pi$.\footnote{$C(c_F)$ has a non analyticity at 
$c_F=0$ if $\rho(\pi) \neq 0$.}
This is the case in the lower gap 
and in particular in the high temperature two gap phases (see \cite{Jain:2013py}). 
We restrict attention to these phases in the rest of this Appendix

\subsection{Dualization of the fermionic gap equation}

The fermionic theory described above has been conjectured
\cite{Aharony:2012nh, Aharony:2012ns} 
(see also \cite{Giombi:2011kc, Maldacena:2011jn,
Maldacena:2012sf, Jain:2013py,GurAri:2012is,Takimi:2013zca})
to be dual to a theory of critical fundamental bosons coupled to 
Chern Simons theory, deformed by the addition of a mass $m_B^{\text{cri}}$ subject to the 
parameter identifications 
\be\label{dict}
\begin{split}
\lambda_F & = \lambda_B- {\rm \sgn}(\lambda_B) \\
 \lambda_B&= \lambda_F-{\rm \sgn}(\lambda_F)\\
 m_{F}^{\text{reg}}&=-\lambda_B m_B\\
 \lambda_F \rho_F(\alpha)&= \lambda_B \rho_B(\alpha+\pi) 
 -\frac{{\rm \sgn}(\lambda_B)}{2\pi} \\
\lambda_B \rho_B(\alpha)&= \lambda_F \rho_F(\alpha+\pi) 
-\frac{{\rm \sgn}(\lambda_F)}{2\pi} \\
c'_{F}&=c_{B}\\
\nu'_{F}&=\nu_{B}.
\end{split}
\ee

\eqref{gap} may be recast in terms of the new variables as\footnote{Notice that the integral in this equation is an even function of ${c_{B}}$. 
It follows that if we set ${c_{B}}={\rm \sgn}(\lambda_B) d$ then the 
equation for $d$ depends only on $|\lambda_B|$ and not on 
the sign of  $\lambda_B$. This had to be the case on physical 
grounds. In the bosonic theory the sign of $\lambda_B$ is unphysical as it 
is flipped by a parity transformation. }
\footnote{Let us pause to analyse the smoothness of the RHS of \eqref{bosegap}. As 
we have explained above, the integral on the RHS of \eqref{bosegap} has 
a non analyticity. This non analyticity cancels that of the second term on 
the RHS of \eqref{bosegap} if and only if $\rho_B(\alpha)=\frac{1}
{2 \pi |\lambda_B|}$ in a finite neighborhood around $\alpha=0$ . This 
is always the case when the original Fermionic theory was in the lower gap 
or two gap phase, as we have assumed in this appendix. In other words
\eqref{bosegap} is perfectly analytic. } 
\begin{equation} \label{bosegap}
\begin{split} 
{c_{B}}&=- \lambda_B \hat m_{B}^{\text{cri}} -{\rm \sgn} (\lambda_B) 
{\rm max}\left( |{c_{B}}|, |\nu_{B}| \right) \\
& +\lambda_B \int_{-\pi}^{\pi} d\alpha~\rho_B(\alpha +\pi)\left(\ln 2\cosh(\frac{c_{B}+i \alpha + \nu_{B}}{2})+\ln 2\cosh(\frac{ c_{B}-i \alpha -\nu_{B} }{2})\right). \\
\end{split}
\end{equation}

\begin{figure}[tbp] 
  \begin{center}
  \subfigure[]{\includegraphics[scale=.4]{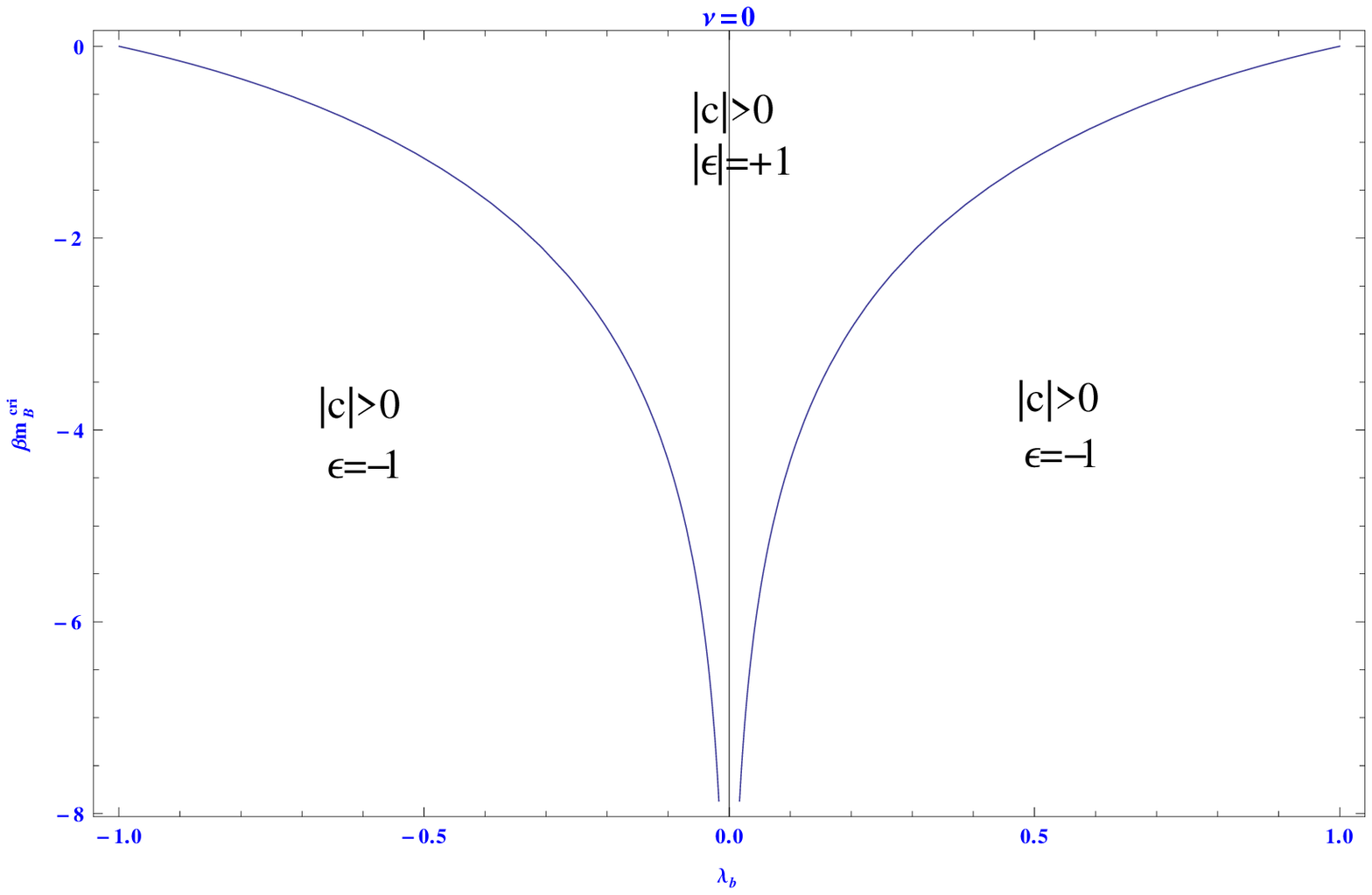}
\label{nu-zero-1}
  }
  \qquad\qquad
  \subfigure[]{\includegraphics[scale=.4]{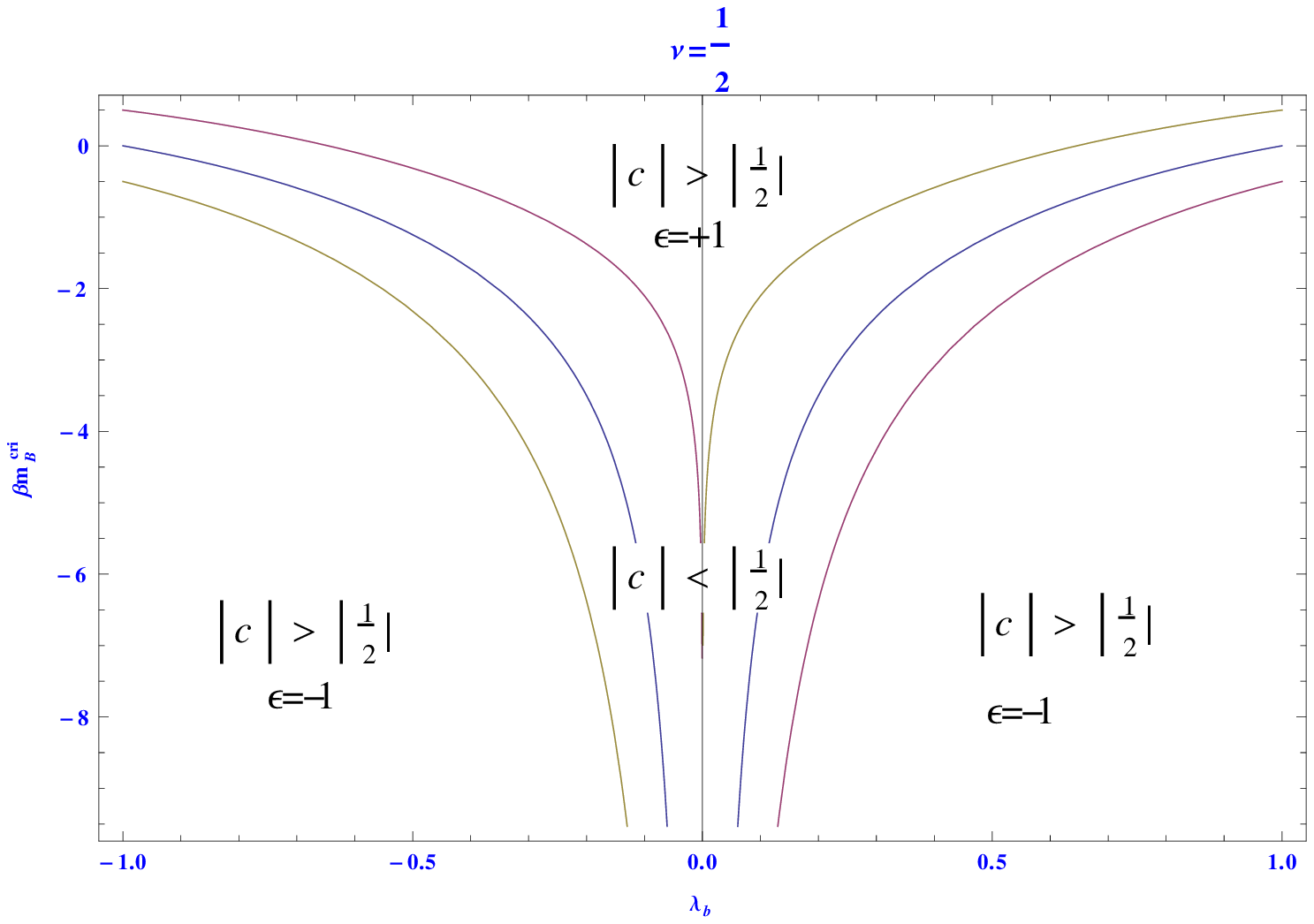}
\label{nu-half}
  }
  \qquad\qquad
 \subfigure[]{\includegraphics[scale=.5]{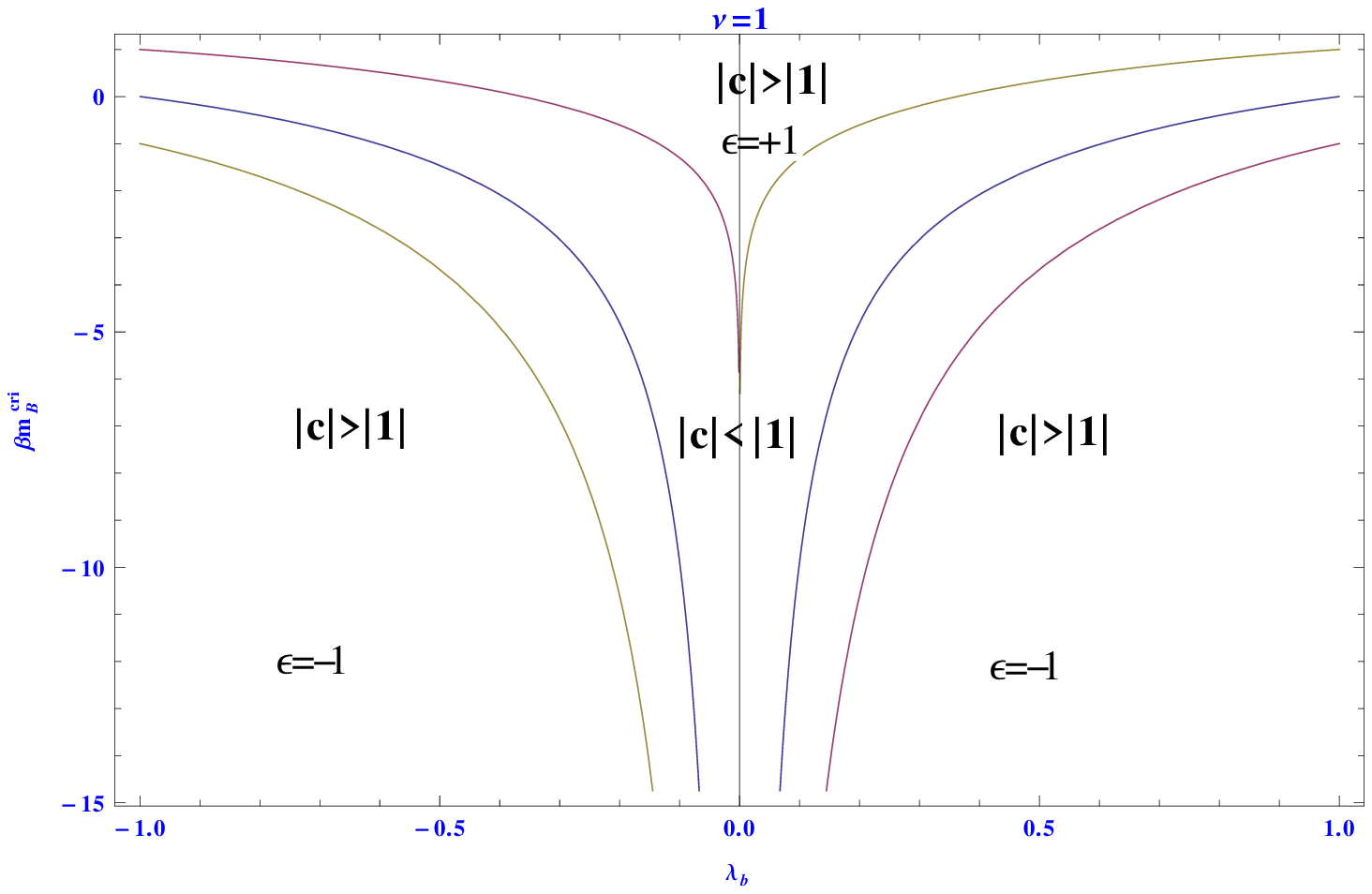}
\label{nu-one}
  }
  \end{center}
  \vspace{-0.5cm}
  \caption{In Fig \ref{nu-zero-1},\ref{nu-half},\ref{nu-one} we indicate where each of the two conditions in \eqref{condition1} 
is obeyed for three separate values of $\nu_B$, namely $|\nu_B|=0, \half, 1$ respectively. The $y$ axis in each plot is $\hat m_{B}^{\text{cri}}$ while the 
$x$ axis is $\lambda_B$. In each of these plots $|c_B|>|\nu_B|$ if we lie either above the highest curve on the graph or below the lowest 
curve on the graph. On the other hand $\epsilon>0$ above the middle (blue) curve, but $\epsilon < 0$ below the blue curve. 
Both conditions in \eqref{condition1} are only met above the highest curve on each plot. }
  \end{figure}

\begin{figure}[tbp]
  \begin{center}
    \subfigure[]{\includegraphics[scale=.5]{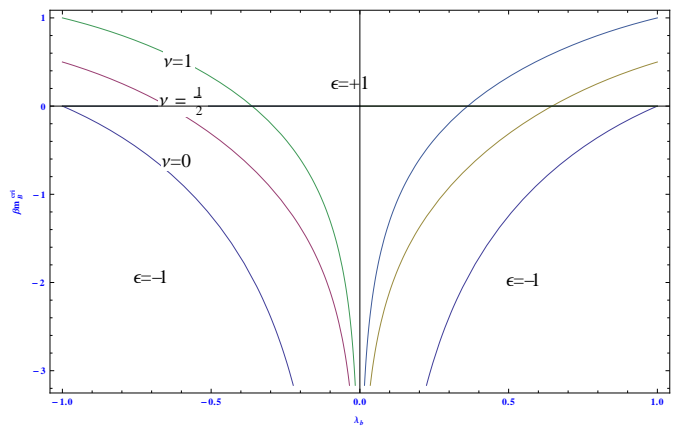}
\label{compare}
}

  \end{center}
  \vspace{-0.5cm}
  \caption{ We replot the highest curve in each of the graphs \ref{nu-zero-1},\ref{nu-half},\ref{nu-one} for comparison. 
The highest curve on this plot is for $|\nu_F|=1$, the middle curve is at $|\nu_F|=\half$ while the lowest curve is 
at $|\nu_F|=0$.}
  \end{figure}

The equation \eqref{bosegap} simplifies when 
\begin{equation}\label{cond}
{c_{B}} + {\rm \sgn}(\lambda_B) {\rm max}(|{c_{B}|}, |\nu_{B}|)=0.
\end{equation}
When this is the case the gap equation reduces to 
\begin{equation} \label{bosegap1}
 \lambda_B \hat m_B^{\text{cri}}= 
\int_{-\pi}^{\pi} d\alpha~\rho_B(\alpha +\pi)\left(\ln 2\cosh(\frac{c_{B}+i \alpha + \nu_{B}}{2})+\ln 2\cosh(\frac{c_{B}-i \alpha -\nu_{B} }{2})\right).
\end{equation}
This is precisely the gap equation for the massive critical boson theory 
in the uncondensed phase. In other words the fermionic gap equation (which 
we believe is accurate at all values of parameters) reduces under duality
 to the gap 
equation of the massive critical boson in the uncondensed phase if and only 
if \eqref{cond} is obeyed. 
This result suggests that the boson lies in a condensed phase 
whenever \eqref{cond} is violated. 

\eqref{cond} is obeyed provided 
\be\label{condition1}
| c_{B}|>|\nu_{B}|,~~\epsilon = -\sgn( c_{B})\sgn(\lm_B)>0.
\ee
For what values of the bosonic Lagrangian parameters
does this simplification happen? In order to answer this question we need to 
solve for $c_B$ as a function of Lagrangian parameters and then constrain 
these parameters using the inequalities above. We have implemented this 
procedure numerically on Mathematica at three different values of the chemical 
potential, $\nu_B=0$, $\nu_B=\frac{1}{2}$ and $\nu_B=1$. For each value of the chemical potential 
we have analyzed when \eqref{cond} is obeyed on a graph whose $y$ axis 
is $\hat m_{B}^{\text{cri}}$ and whose $x$ axis is $\lambda_B$ in Fig \ref{nu-zero-1},\ref{nu-half},\ref{nu-one}.

On intuitive grounds we expect the scalar to show an increasing tendency 
to condense at larger values of  $|\nu_B|$ and at increasingly 
negative values of $m_{B}^{\text{cri}}$ (when the scalar field theory is increasingly 
tachyonic). We also expect finite temperature condensation to be forbidden 
by the Mermin Wagner theorem at $\lambda_B=0$. Our explicit results 
are in perfect agreement with this intuition. We find that the 
conditions listed in \eqref{condition1} are met provided we lie `above' the highest curves 
in this plots given in Fig \ref{nu-zero-1},\ref{nu-half},\ref{nu-one}.% 
\footnote{In each of these plots the first condition in \eqref{condition1}, ($|c_{B}|>|\nu_{B}|$) is met if 
we lie either above the highest curves on the plot or below the lowest curves on that plot. However 
the second condition in \eqref{condition1}, namely $\epsilon>0$, is met only above the highest 
curves.} Note that, on each of these plots,  increasingly negative values 
of $m_{B}^{\text{cri}}$ and increasingly large values of $|\nu_B|$ tend to cause condensation. 
Note also that condensation becomes increasingly unlikely at small $\lambda_B$ 
and never occurs at $\lambda_B=0$. 
In Fig. \ref{compare} have replotted the highest curves of Figs. 
 Fig \ref{nu-zero-1},\ref{nu-half},\ref{nu-one} on the same graph for comparison. 
Note that the curve rises as $|\nu_B|$ is increased. 
This observation supports the
intuition that increasing $|\nu_B|$ increases the probability of 
``condensation''. All in all our results qualitatively support the guess that 
the bosonic field Bose condenses when either of the inequalities in 
\eqref{condition1} are violated. 

\subsection{Dualization of the bosonic gap equation}
In this subsection we demonstrate that no solution exists to the 
critical boson theory when the (dual of) the inequality 
$\sgn(\lambda) c_F>0$ is violated. 

The gap equation of the critical boson theory is 
\be\begin{split}\label{bee}
2\lm_{B}\cS &=\lm_B m_B^{\text{cri}}.\\
\end{split}
\ee
We can rewrite this equation in fermionic dual variables as 
\be
\sgn(\lm_F){\rm max}\left( |{c_{F}}|, |\nu_{F}| \right) =\left(2\lm_F \cC+m_{F}^{\text{reg}}\right).
\ee
By comparing the signs of the RHS and LHS of this equation, it follows 
immediately that the equation has no solution unless 
\be \label{ineqa}
\sgn \(\lm_F\left(2\lm_F \cC+m_{F}^{\text{reg}}\right)\)>0
\ee
as we set out to show. 

When this inequality is obeyed and  
$|{c_{F}}|> |\nu_{F}|.$ it follows that 
\be\begin{split}
\sgn(\lm_F)|{c_{F}}| &=\left(2\lm_F \cC+m_{F}^{\text{reg}}\right)\\
|{c_{F}}| &=\sgn(\lm_F) \left(2\lm_F \cC+m_{F}^{\text{reg}}\right).
\end{split}\ee
This is the fermionic gap equation. Note that \eqref{bee} may have solutions
when \eqref{ineqa} is obeyed but $c_B<\nu_B$.

\bibliographystyle{JHEP}
\bibliography{csmd}

\end{document}